\documentclass[journal=mamobx,manuscript=article]{achemso}

\usepackage{chemformula} 
\usepackage[T1]{fontenc} 
\usepackage{amssymb}
\usepackage{mathrsfs}
\usepackage{setspace}
\usepackage{natbib}

\newcommand{\fref}[1]{Figure~\ref{#1}}
\newcommand{\sref}[1]{Section~\ref{#1}}

\SectionNumbersOn

\author{M. Manav}
\affiliation{Graduate Aerospace Laboratories, California Institute of Technology, Pasadena, USA}
\email{manav@alumni.ubc.ca}
\author{M. Ponga}
\affiliation{Mechanical Engineering, University of British Columbia, Vancouver, Canada}
\email{mponga@mech.ubc.ca}
\author{A. Srikantha Phani}
\affiliation{Mechanical Engineering, University of British Columbia, Vancouver, Canada}
\email{srikanth@mech.ubc.ca}

\title{Stress in a stimuli-responsive polymer brush}

\begin{document}

\begin{singlespace}
	
\begin{abstract}
	The application of a polymer brush in sensing, actuation, self-folding, among others acutely depends on the tuneable bending of a brush-grafted substrate caused by the stress in the brush. However, the stress in a stimuli-responsive brush has not been investigated. In this work, we study the stress in the stimuli-responsive planar polymer brushes of neutral water-soluble polymers with low to very high graft densities using strong stretching theory (SST). First, SST with the Langevin force-extension relation for a polymer chain is extended to the study of stimuli-responsive brushes. Stress profile and other properties of a Poly(N-isopropylacrylamide) (PNIPAm) brush are then obtained using the extended SST and an empirical Flory-Huggins parameter. The model predicts that the stress in a PNIPAm brush is inhomogeneous and compressive at all temperatures and graft densities. The resultant stress is predicted to increase in magnitude with increasing graft density. Moreover, it decreases in magnitude with an increase in temperature before plateauing in low graft density brushes. In contrast, its magnitude increases weakly with increasing temperature in high density brushes. This contrasting behavior is traced to the minimum in interaction free energy density \emph{vs} polymer volume fraction curve for PNIPAm solution at a large volume fraction, and stiffening of chains due to finite extensibility. Furthermore, our results indicate that the ability to tune the resultant stress by changing temperature diminishes with increasing graft density.
\end{abstract}

\section{Introduction}
Polymer chains densely grafted to an impermeable substrate interact with each other and stretch away from the grafting surface to form a polymer brush\cite{milner91,alexander77,de80,milner88,halperin94,netz03,binder12,azzaroni2018}. The structure of a brush arises from a balance between the excluded volume interaction among monomers and the elastic stretching of the polymer chains with an end-grafting constraint. Brushes have promising applications in surface lubrication, antifouling surfaces to mention but a few \cite{stuart2010,azzaroni2012,chen2017,azzaroni2018}. A \emph{key} property of a polymer brush is the existence of a normal stress parallel to the substrate, resulting from the balance between chain stretching and the interaction among monomers, and tunable by stimuli. When grafted onto a flexible substrate, this tuneable stress can bend the substrate allowing the application of brushes in sensing\cite{abu2006micro,klushin2014,chen2010,peng2017}, actuation\cite{zhou2006,zhou2008,zou2011}, self-folding\cite{kelby2011,qi2016,xu2017,xu2018}, among others.

Recent years have witnessed a spurt in research on the applications leveraging brush induced tunable substrate deformation.\cite{abu2006micro,klushin2014,chen2010,peng2017,zhou2006,zhou2008,zou2011,kelby2011,qi2016,xu2017,xu2018} Polymer brushes have been shown to bring about shape changes in two dimensional (2D) materials to achieve compact form factor which can benefit applications such as wearable electronics, biological sensors, actuators, to name a few.\cite{xu2017,xu2018} Shape change in these materials could also modify the physical and chemical properties enabling novel applications such as in energy storage, strain sensing etc.\cite{deng2016} The key to control and manipulate substrate deformation depends on a fundamental understanding of the stress within a brush, and its variation upon the application of stimuli. However, this aspect of a brush has not been fully addressed. Xu~\emph{et al.}\cite{xu2017} used molecular dynamics (MD) simulations to calculate PNIPAm brush mechanical properties including stress. Unfortunately, MD simulations are computationally expensive. Manav~\emph{et al.}\cite{manav2018} used strong stretching theory (SST)\cite{milner88,skvortsov88,semenov85} to study the stress variation within a brush of Gaussian chains. Their theory predicts that the stress is compressive and varies as a quartic function of distance from the grafting surface in a good solvent. An extension to high density brushes\cite{shim89,amoskov94,biesheuvel2008} using semi-analytical SST with the Langevin force-extension relation for a chain was latter presented\cite{manav2019}. This work also validated the SST results on the stress variation within brushes of low to very high graft densities using molecular dynamics simulation. It should be mentioned here that the inhomogeneous stress within a brush has never been measured experimentally. Earlier, the resultant stress and the elastic modulus of the brush grafted on an impermeable substrate were obtained as a function of graft density, length of a chain, and solvent strength using scaling theory\cite{utz08}. Begley~\emph{et al.}\cite{begley2005} studied the deformation in a beam due to a grafted brush layer by simplifying the brush layer as interacting beads on the surface of the beam. A model for the large deformation of a brush grafted beam was developed by Manav~\emph{et al.}\cite{manav2018} incorporating  curvature elasticity and the Young-Laplace corrections to the classical Stoney's equation\cite{stoney09}.

The effect of stimuli on a moderately dense brush of a classical polymer\cite{flory53,de79} has been studied extensively by Zhulina \emph{et al.}\cite{zhulina91} using both scaling theory\cite{alexander77,de80,halperin94} and SST\cite{milner88,skvortsov88}. They have shown that a brush is inhomogeneous at any solvent strength and obtained analytical expressions for polymer volume fraction and end density profiles. The volume fraction was predicted to jump abruptly at the brush free end in a poor solvent instead of smoothly going to zero. A planar brush of a classical polymer was shown to undergo a cooperative transition instead of a true thermodynamic phase transition due to a change in solvent quality\cite{zhulina91,birshtein1997}. Shim and Cates\cite{shim89} reported change in monomer density profile of a high density brush of a classical polymer as solvent quality is varied. To model a brush of non-classical polymers such as Poly(ethylene oxide) (PEO) and PNIPAm, a two-state model for the solutions of such polymers\cite{karlstrom1985,matsuyama1990,bekiranov1997,baulin2002} was extended to brushes by Baulin \emph{et al.}\cite{baulin2003self}. This model used a Flory-Huggins parameter that depends on both temperature as well as monomer density.\cite{baulin2003self,mendez2005} They elucidated the possibility of a discontinuous polymer volume fraction profile due to vertical phase separation within the brush, which has been observed in PNIPAm brushes experimentally\cite{yim2005,laloyaux2009,varma2016} as well. Halperin \emph{et al.}\cite{halperin2011} utilized this model to study the collapse of a thermoresponsive brush as a consequence of a change in temperature and the resulting change in protein adsorption in the brush. The applicability of the work of Baulin \emph{et al.}\cite{baulin2003self}, however, is limited to low to moderately dense brushes only because it assumes brushes to be made of Gaussian chains.

The effect of stimuli on the stress distribution in a brush has never been explicitly investigated. Furthermore, the effect of stimuli on high density brushes remains largely unexplored. This work attempts to address these gaps in polymer brush literature using SST with the Langevin force-extension relation for a polymer chain. We first obtain semi-analytical expressions for the end density profile within a brush of any graft density at an arbitrary solvent strength. Together with the polymer volume fraction within a brush, they are then used to calculate the stress within the brush. The two key contributions of this paper are (a) calculation of the structural and other properties of a high graft density stimuli-responsive brush, and (b) the stress distribution in a stimuli-responsive brush. To show the effect of stimuli on a brush, in the numerical calculations, we chose Flory-Huggins parameter for PNIPAm reported by Afroze \emph{et al.}\cite{afroze2000}. The choice of PNIPAm is guided by frequent use of PNIPAm brushes in applications involving brushes. Polymer volume fraction, end density, stress and other properties will be evaluated and discussed for these brushes.

Rest of the paper is organized as follows. \sref{th} presents an extension of the SST framework for calculation of polymer volume fraction and end density profile of a brush of any graft density and at any solvent strength. Calculation of stress within the SST framework in a stimuli-responsive brush is also discussed in this section. Results on the effect of temperature, graft density and chain length on PNIPAm brush properties is presented in \sref{stimuliEffect} and discussed in \sref{discussion}. The paper ends with conclusions in \sref{conclusion}.

\section{SST for stimuli-responsive brushes}
\label{th}
\begin{figure}[!h]
	\center
	\includegraphics[width=8cm]{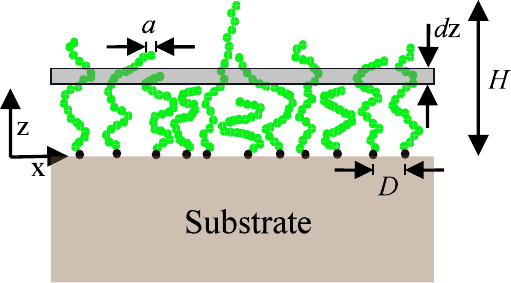}
	\caption{The schematic shows a planar polymer brush of height $H$, made of the polymer chains with effective monomer size $a$. Average distance between the grafting points in the brush equals inverse square root of graft density ($\langle D\rangle=\rho_g^{-1/2}$). A thin layer of infinitesimal thickness $d\textsc{z}$ at height $\textsc{z}$ is also shown. Grafted monomers are shown in black color. Also, $\textsc{y}$-axis goes into the plane following right-hand rule.}
	\label{brush_slit}
\end{figure}

Consider a planar monodisperse brush (see \fref{brush_slit}) of graft density $\rho_g$, in which each chain has $N$ effective monomers of size $a$ and volume $v_0$. The Kuhn length of a polymer chain is $l_k=r_ka$, where $r_k$ is asymmetry parameter. The height of the brush is denoted by $H$. The polymer volume fraction $\phi$ within a brush is homogeneous in any plane parallel to the grafting surface. The total free energy of the brush is given by:
\begin{equation}
F=F_{int}+F_{el}+F_{interface}+F_{substrate},
\label{eq0}
\end{equation}
where $F_{int}$ is interaction free energy arising from the interaction among monomers in the brush, and $F_{el}$ is the free energy associated with the elastic stretching of the polymer chains in the brush. $F_{interface}$ is the free energy associated with the surface and interfaces in a brush. This term accounts for the contributions made by the interface between the brush free end and pure solvent, the brush-substrate interface, and the interface between the two phases within the brush in case of a phase separation. $F_{substrate}$ is strain energy of the substrate on which the brush is grafted. Since the effect of the surface and interfaces is localized in the vicinity of the interface in a brush with long polymer chains, we substitute $F_{interface}=0$ and ignore their effect in this work. Also, because substrate deformation influences brush properties primarily by changing graft density as long as the brush with the deformed substrate can be assumed to be planar, it is inconsequential if the substrate is rigid or flexible (in equilibrium). Hence, we assume substrate to be rigid and ignore $F_{substrate}$ to simplify the problem. Note that a brush with a deformed substrate can be assumed to be planar if the characteristic length scale of substrate deformation (such as the radius of curvature in bending) is much larger than the brush height. Under these assumptions, the free energy of the brush given in \eqref{eq0} simplifies to:
\begin{equation}
F=F_{int}+F_{el}.
\label{eq01}
\end{equation}
The structure of a brush is obtained by minimizing this free energy \cite{zhulina91}. Within a mean field approximation, a brush can also be described using a polymer chain in a one-dimensional self-consistent field. When the polymer chains in a brush are strongly stretched \cite{semenov85,milner88,skvortsov88,netz98}, a chain can be described as a spring in a nonuniform stretching potential field\cite{de79,amoskov94} $\bar{V}(\bar{\textsc{z}})$, where  $\bar{\textsc{z}}=\textsc{z}/Na$. The chemical potential $\bar{\mu}(\bar{\phi}(\bar{\textsc{z}}))$, where $\bar{\phi}(\bar{\textsc{z}})=\phi(\textsc{z})$ is polymer volume fraction, in the brush provides the potential field $\bar{V}(\bar{\textsc{z}})$ such that
\begin{equation}
\bar{V}(\bar{H})-\bar{V}(\bar{\textsc{z}})=\bar{\mu}(\bar{\phi}(\bar{\textsc{z}}))-\bar{\mu}(\bar{\phi}(\bar{H})), \label{eq1}
\end{equation}
where $\bar{H}=H/Na$. Additionally, the local stretching force $\bar{p}$ at $\bar{\textsc{z}}$ in a chain with the end at $\bar{\textsc{z}}_e$ is related to the potential field by the following relation:\cite{amoskov94}
\begin{equation}
\bar{E}(\bar{p})=\frac{1}{r_k}\int_0^{\bar{p}} \bar{e}(\bar{p}')d\bar{p}'=\bar{V}(\bar{\textsc{z}}_e)-\bar{V}(\bar{\textsc{z}}), \label{eq5b}
\end{equation}
where, $\bar{E}(\bar{p})$ is complementary stretching energy, and $\bar{e}(\bar{p}')$ is the local stretching in a chain due to a stretching force $\bar{p}'$ and is given by the force-extension relation for a polymer chain. The constraint that each chain has $N$ monomers, and the self consistency condition for a brush yield the following equations:\cite{amoskov94}
\begin{align}
&\int_0^{\bar{\textsc{z}}_e}\frac{d\bar{\textsc{z}}}{ \bar{\Lambda}(\bar{V}(\bar{\textsc{z}}_e)-\bar{V}(\bar{\textsc{z}}))}=1,\label{eq7b} \\
&\bar{\phi}(\bar{\textsc{z}})=\int_{\bar{\textsc{z}}}^{\bar{H}}\frac{\bar{g}(\bar{\textsc{z}}_e)}{\bar{\Lambda}(\bar{V}(\bar{\textsc{z}}_e)-\bar{V}(\bar{\textsc{z}}))}d\bar{\textsc{z}}_e, \label{eq9a}
\end{align}
where $\bar{g}(\bar{\textsc{z}}_e)$ is end density at $\bar{\textsc{z}}_e$. Also, $\bar{\Lambda}(\bar{V}(\bar{\textsc{z}}_e)-\bar{V}(\bar{\textsc{z}}))=\bar{e}(\bar{p})$, where $\bar{p}$ is the local stretching force at $\bar{\textsc{z}}$ in a chain with the end at $\bar{\textsc{z}}_e$. Appendix~\ref{SST_formulation} summarizes the derivation of the above equations following Amoskov \emph{et al.}\cite{amoskov94} and also describes the nondimensionalization scheme.

For a finitely extensible freely-jointed ideal chain model, used in this work, the force-extension relation is given by Langevin function ($\mathcal{L}(.)$).
\begin{equation}
\bar{e}(\bar{p})=\mathcal{L}(\bar{p})=\coth (\bar{p})-\frac{1}{\bar{p}}, \quad {\rm and} \quad \bar{E}(\bar{p})=\frac{1}{r_k}\ln\left(\frac{\sinh(\bar{p})}{\bar{p}}\right). \label{eq9b}
\end{equation}
By solving \eqref{eq7b} for this chain model, a series solution for $\bar{V}(\bar{\textsc{z}})$ was obtained\cite{amoskov94}. A rational polynomial approximation for the same was also reported\cite{biesheuvel2008} as:
\begin{equation}
\bar{V}(\bar{\textsc{z}})=\frac{2\bar{\textsc{z}}^2}{r_k}\left(\frac{2-\frac{4}{5}\bar{\textsc{z}}^2}{1-\bar{\textsc{z}}^2}\right). \label{eq8}
\end{equation}
This approximation along with an appropriate choice of chemical potential has been shown to yield brush properties in a good solvent which quantitatively agree with the results obtained from molecular dynamics simulation\cite{biesheuvel2008,manav2019}. Since \eqref{eq7b} is independent of solvent strength and only depends on the form of $\bar{e}(\bar{p})$, we use the above expression for $\bar{V}(\bar{\textsc{z}})$ in this work as well.

Once $\bar{V}(\bar{\textsc{z}})$ is known, $\bar{\phi}(\bar{\textsc{z}})$ can be obtained using \eqref{eq1}. Then, \eqref{eq9a} can be solved to obtain $\bar{g}(\bar{\textsc{z}})$. This has been done semi-analytically for high density swollen polymer brushes\cite{amoskov94,biesheuvel2008} only. In this work, we solve Eq. \eqref{eq9a} to obtain the end density in a stimuli-responsive brush by taking into consideration the possible discontinuity in polymer volume fraction profile induced by stimuli. Two distinct cases have been considered in the derivation. In the first case, no vertical phase separation occurs within the brush and volume fraction profile is continuous, though a jump in the profile at the brush free end, observed in a collapsed brush, can occur. In this case, the end density is:
\begin{equation}
\bar{g}(\bar{\textsc{z}})=\frac{d\bar{V}(\bar{\textsc{z}})}{d\bar{\textsc{z}}}\left(\int_0^{\bar{\textsc{z}}_f}\frac{dR(\bar{u}-\bar{v})}{d(\bar{u}-\bar{v})}d\bar{\textsc{z}}'_f+R(0)\frac{d\bar{\textsc{z}}_f}{d\bar{u}} \right), \label{eq10}
\end{equation}
where,
\begin{equation}
\bar{u}=\bar{V}(\bar{H})-\bar{V}(\bar{\textsc{z}})=\bar{V}(\bar{\textsc{z}}_f), \quad \bar{v}=\bar{V}(\bar{H})-\bar{V}(\bar{\textsc{z}}_e)=\bar{V}(\bar{\textsc{z}}'_f), \quad R(\bar{u})=\bar{\phi}(\bar{\textsc{z}}), \label{eq11}
\end{equation}
and $R(0)=\bar{\phi}(\bar{H})$. Hence, the last term in the expression for $\bar{g}(\bar{\textsc{z}})$ contributes only if there is a discontinuity in the polymer volume fraction profile at the free end of the brush like in a collapsed brush. Also, in that case, the end density has a vertical asymptote at the free end of brush.

In the second case, a brush with a vertical phase separation inside the brush is considered. For a brush with vertical phase separation at height $\bar{H}_t$, the end density is given by the following expression:
\begin{equation}
\bar{g}(\bar{\textsc{z}})=\begin{cases}
\frac{d\bar{V}(\bar{\textsc{z}})}{d\bar{\textsc{z}}}\left(\int_0^{\bar{\textsc{z}}_f}\frac{dR(\bar{u}-\bar{v})}{d(\bar{u}-\bar{v})}d\bar{\textsc{z}}'_f+R(0)\frac{d\bar{\textsc{z}}_f}{d\bar{u}} \right), & \bar{\textsc{z}}\ge\bar{H}_t\\
\frac{d\bar{V}(\bar{\textsc{z}})}{d\bar{\textsc{z}}}\left(\int_0^{\bar{\textsc{z}}_g}\frac{dS(\tilde{u}-\tilde{v})}{d(\tilde{u}-\tilde{v})}d\bar{\textsc{z}}'_g+S(0)\frac{d\bar{\textsc{z}}_g}{d\tilde{u}}\right), & \bar{\textsc{z}}<\bar{H}_t
\end{cases}
\label{eq12}
\end{equation}
where, for $\bar{\textsc{z}}\ge\bar{H}_t$, expression is the same as in \eqref{eq10}. Also,  $\tilde{u}=\bar{V}(\bar{H}_t)-\bar{V}(\bar{\textsc{z}})=\bar{V}(\bar{\textsc{z}}_g)$, $\tilde{v}=\bar{V}(\bar{H}_t)-\bar{V}(\bar{\textsc{z}}_e)=\bar{V}(\bar{\textsc{z}}'_g)$, and
\begin{equation}
S(\tilde{u}-\tilde{v})=R(\tilde{u}-\tilde{v}+\alpha)-\int_{\bar{H}_t}^{\bar{H}}\frac{\bar{g}(\bar{\textsc{z}}_h)}{\bar{\Lambda}(\tilde{u}-\tilde{v}+\alpha-\bar{v}_h)}d\bar{\textsc{z}}_h,
\label{eq13}
\end{equation}
where, $\alpha=\bar{V}(\bar{H})-\bar{V}(\bar{H}_t)$ and $\bar{v}_h=\bar{V}(\bar{H})-\bar{V}(\bar{\textsc{z}}_h)$. Note that $S(0)=\bar{\phi}(\bar{H}_t-\delta)-\bar{\phi}(\bar{H}_t+\delta)$, $\delta\to 0$, that is, it equals the jump in the polymer volume fraction at $\bar{\textsc{z}}=\bar{H_t}$. Also, since $\bar{\textsc{z}}_g \to 0$ and $\frac{d\tilde{u}}{d\bar{\textsc{z}}_g}\to 0$ when $\bar{\textsc{z}} \to \bar{H}_t$, the end density diverges and has a vertical asymptote at $\bar{\textsc{z}}=\bar{H}_t$ within the brush. See Appendix~\ref{calc_endden} for the derivation of the expressions for $\bar{g}(\bar{\textsc{z}})$ in \eqref{eq10} and \eqref{eq12}, and their calculation.

\subsection{Free energy density and stress}
\label{stress}
Within SST framework, the free energy density at any height within a brush ($f(\bar{\textsc{z}})$) can be calculated. It consists of two parts: interaction free energy density ($f_{int}(\bar{\textsc{z}})$) and stretching free energy density ($f_{el}(\bar{\textsc{z}})$). Using Flory-Huggins solution theory\cite{flory1942thermodynamics}, $f_{int}(\bar{\textsc{z}})$ for a brush of long polymer chains is expressed as:
\begin{equation}
f_{int}(\bar{\textsc{z}})=\left[(1-\bar{\phi}(\bar{\textsc{z}}))\ln(1-\bar{\phi}(\bar{\textsc{z}}))+\chi \bar{\phi}(\bar{\textsc{z}})(1-\bar{\phi}(\bar{\textsc{z}}))\right]\frac{k_BT}{v_0}, \label{eq14}
\end{equation}
where $\chi$ is the Flory-Huggins interaction parameter. The stretching free energy density at height $\bar{\textsc{z}}$ equals \cite{manav2019}:
\begin{equation}
f_{el}(\bar{\textsc{z}})=\int_{\textsc{z}}^H\frac{dF_{chain}}{d\textsc{z}}g(\textsc{z}_e)d{\textsc{z}_e}=\int_{\bar{\textsc{z}}}^{\bar{H}}\frac{a}{v_0}\frac{dF_{chain}}{d\textsc{z}}\bar{g}(\bar{\textsc{z}}_e)d{\bar{\textsc{z}}_e}, \label{eq15}
\end{equation}
where $\frac{dF_{chain}}{d\textsc{z}}$ is the free energy of stretching per unit length of a part of single chain inside the thin layer in \fref{brush_slit} whose end is at height $\textsc{z}_e$. For a chain with the Langevin force-extension relation \cite{manav2019},
\begin{equation}
\frac{dF_{chain}}{d\textsc{z}}=\frac{k_BT}{r_ka\bar{e}(\bar{p})}\left(\bar{p}\bar{e}(\bar{p})-\ln\left(\frac{\sinh(\bar{p})}{\bar{p}} \right)\right). \label{eq16}
\end{equation}

To obtain the stress in a brush, an infinitesimally small uniform horizontal strain $\delta\epsilon_{\textsc{xx}}$ is applied to the brush. The applied strain changes the free energy within a thin layer at height $\bar{\textsc{z}}$ shown in \fref{brush_slit}. The change in the free energy equals the change in the strain energy within the layer. Assuming all the shear stresses as well as the normal stress in the $\textsc{z}$-direction to be zero and plane strain condition in the $\textsc{y}$-direction ($\epsilon_{\textsc{yy}}=0$),
\begin{equation}
\delta(f(\bar{\textsc{z}})V)=\sigma_{\textsc{xx}}\delta\epsilon_{\textsc{xx}}V_0,
\label{eq17a}
\end{equation}
where $V_0$ is the initial volume of the layer. 

A complete characterization of the stress state within a brush requires us to consider two possible cases: first, the stress in the sections of a brush with continuous volume fraction, and second, the stress at the interface in case of a vertical phase separation within the brush. In the first case\cite{manav2018,manav2019}, the above expression simplifies to:
\begin{equation}
\sigma_{\textsc{xx}}\delta\epsilon_{\textsc{xx}}=\delta f(\bar{\textsc{z}})+f(\bar{\textsc{z}})\frac{\delta V}{V_0},
\label{eq17b}
\end{equation}
where, $\frac{\delta V}{V_0}$ is the volumetric strain arising in the layer due to the applied $\delta\epsilon_{\textsc{xx}}$. It equals $(\delta\epsilon_{\textsc{xx}}+\delta\epsilon_{\textsc{yy}}+\delta\epsilon_{\textsc{zz}})=(\delta\epsilon_{\textsc{xx}}+\delta\epsilon_{\textsc{zz}})$, where $\delta\epsilon_{\textsc{zz}}$ is the strain induced in the $\textsc{z}$-direction in a layer at $\bar{\textsc{z}}$ shown in \fref{brush_slit}. On substituting this in Eq. \eqref{eq17b} and taking the limit $\delta\epsilon_{\textsc{xx}}\to 0$, we obtain the stress distribution\cite{manav2018} in the sections of a brush with continuous volume fraction profile.
\begin{eqnarray}
\sigma_{\textsc{xx}}(\bar{\textsc{z}})=\frac{\partial f(\bar{\textsc{z}})}{\partial\epsilon_{\textsc{xx}}}+f(\bar{\textsc{z}})\left(1+\frac{\partial\epsilon_{\textsc{zz}}(\bar{\textsc{z}})}{\partial\epsilon_{\textsc{xx}}}\right),
\label{eq18}
\end{eqnarray}
The first term in the above expression for stress denotes the change in the free energy density due to the applied strain, and the second term relates to the change in the volume of the thin layer due to the applied strain. The derivation of the expression and the calculation of the different terms in \eqref{eq18} follows the approach reported in Manav \emph{et al.}\cite{manav2019} for a brush in a good solvent and the associated calculations are described in Appendix~\ref{stress_calc}.

\begin{figure}[!h]
	\center
	\includegraphics[width=8cm]{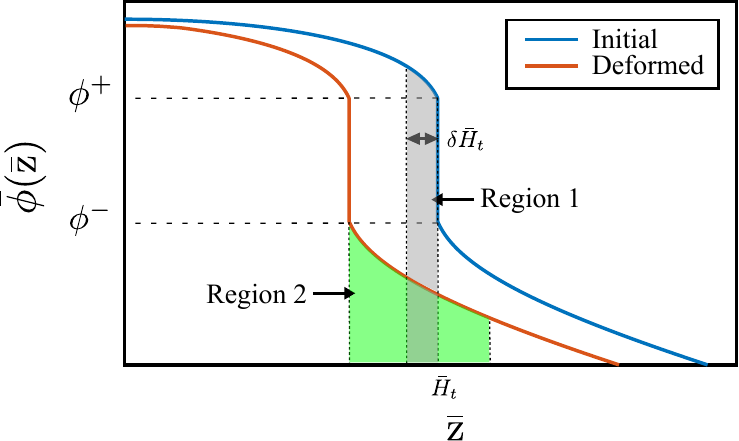}
	\caption{Schematic shows the polymer volume fraction ($\bar{\phi}(\bar{\textsc{z}})$) \emph{vs} distance from the grafting surface ($\bar{\textsc{z}}$) profile in the initial configuration and in the deformed configuration after the strain $\delta\epsilon_{\textsc{xx}}$ is applied. Due to the applied strain, monomers in region 1 (shown by grey filling) move to region 2 (shown by green filling). Also, the left boundary of region 1 becomes the new interface and the left boundary of region 2. Right boundary of region 1 becomes the right boundary of region 2 after deformation.}
	\label{phi_Z_typical}
\end{figure}
In case of a vertical phase separation within the brush, at the interface, $\delta(f(\bar{\textsc{z}})V)$ in \eqref{eq17a} equals the difference in the free energy contained in region 2 of the deformed brush and region 1 of the initial configuration of the brush in \fref{phi_Z_typical}.
\begin{equation}
f_2A(1+\delta\epsilon_{\textsc{xx}})(\delta\bar{H}_t+[\delta\bar{u}_{\textsc{z}}]_{\bar{\textsc{z}}=\bar{H}_t}-[\delta\bar{u}_{\textsc{z}}]_{\bar{\textsc{z}}=\bar{H}_t-\delta\bar{H}_t})-f_1A\delta\bar{H}_t=\sigma_{\textsc{xx}}\delta\epsilon_{\textsc{xx}}A\delta\bar{H}_t,
\label{eq19b}
\end{equation}
where $A$ is substrate surface area, and $\delta\bar{u}_{\textsc{z}}=\delta u_{\textsc{z}}/(Na)$ is the nondimensionalized displacement of the monomers at $\bar{\textsc{z}}$ in the $\textsc{z}$-direction due to the applied strain $\delta\epsilon_{\textsc{xx}}$. $f_2$ is the free energy density in region 2 at right above the interface ($\bar{H}_t+\delta$, $\delta\to 0$) in the deformed brush, and $f_1$ is the free energy density in region 1 at right below the interface ($\bar{H}_t-\delta$, $\delta\to 0$) in the initial brush. $\delta\bar{H}_t$ is chosen such that on applying $\delta\epsilon_{\textsc{xx}}$, the monomers at $\bar{H}_t-\delta\bar{H}_t$ in the undeformed brush lie at the new interface in the deformed brush. $\phi^+$ and $\phi^-$ in \fref{phi_Z_typical} are the coexisting polymer volume fractions at the interface and will be discussed in the next section. On dividing by $(\delta\epsilon_{\textsc{xx}})^2$, \eqref{eq19b} simplifies to the following in the limit of $\delta\epsilon_{\textsc{xx}}\to 0$:
\begin{equation}
\sigma_{\textsc{xx}}\frac{\partial\bar{H}_t}{\partial\epsilon_{\textsc{xx}}}=\lim_{\delta\epsilon_{\textsc{xx}}\to 0}\frac{f_2(1+\delta\epsilon_{\textsc{xx}})\left(\frac{\partial\bar{H}_t}{\partial\epsilon_{\textsc{xx}}}+\left[\frac{\partial\bar{u}_{\textsc{z}}}{\partial\epsilon_{\textsc{xx}}}\right]_{\bar{\textsc{z}}=\bar{H}_t}-\left[\frac{\partial\bar{u}_{\textsc{z}}}{\partial\epsilon_{\textsc{xx}}}\right]_{\bar{\textsc{z}}=\bar{H}_t-\delta\bar{H}_t}\right)-f_1\frac{\partial\bar{H}_t}{\partial\epsilon_{\textsc{xx}}}}{\delta\epsilon_{\textsc{xx}}}.
\label{eq19c}
\end{equation}
In a brush with vertical phase separation, $\frac{\partial\bar{u}_{\textsc{z}}}{\partial\epsilon_{\textsc{xx}}}$ is also discontinuous and $\left[\frac{\partial\bar{u}_{\textsc{z}}}{\partial\epsilon_{\textsc{xx}}}\right]_{\bar{\textsc{z}}=\bar{H}_t}-\left[\frac{\partial\bar{u}_{\textsc{z}}}{\partial\epsilon_{\textsc{xx}}}\right]_{\bar{\textsc{z}}}$ equals the jump in the profile at $\bar{H}_t$ and is finite. To find $\frac{\partial\bar{H}_t}{\partial\epsilon_{\textsc{xx}}}$, we make use of mass conservation, that is the volume of the monomers contained in region 1 and region 2 in \fref{phi_Z_typical} are equal.
\begin{equation}
\phi^-A(1+\delta\epsilon_{\textsc{xx}})(\delta\bar{H}_t+[\delta\bar{u}_{\textsc{z}}]_{\bar{\textsc{z}}=\bar{H}_t}-[\delta\bar{u}_{\textsc{z}}]_{\bar{\textsc{z}}=\bar{H}_t-\delta\bar{H}_t})-\phi^+A\delta\bar{H}_t=0.
\label{eq19d}
\end{equation}
On dividing the above equation by $\delta\epsilon_{\textsc{xx}}$, taking the limit $\delta\epsilon_{\textsc{xx}}\to 0$ and simplifying, we obtain
\begin{equation}
\frac{\partial\bar{H}_t}{\partial\epsilon_{\textsc{xx}}}=\frac{\phi^-}{\phi^+-\phi^-}\left(\left[\frac{\partial\bar{u}_{\textsc{z}}}{\partial\epsilon_{\textsc{xx}}}\right]_{\bar{\textsc{z}}=\bar{H}_t}-\left[\frac{\partial\bar{u}_{\textsc{z}}}{\partial\epsilon_{\textsc{xx}}}\right]_{\bar{\textsc{z}}=\bar{H}_t-\delta\bar{H}_t}\right).
\label{eq19e}
\end{equation}
We substitute this in \eqref{eq19c} to obtain the expression for the stress at the interface.
\begin{equation}
\sigma_{\textsc{xx}}(\bar{H}_t)=\lim_{\delta\epsilon_{\textsc{xx}}\to 0}\frac{1}{\phi^-}\frac{f_2\phi^+-f_1\phi^-}{\delta\epsilon_{\textsc{xx}}}.
\label{eq19f}
\end{equation}
See Appendix~\ref{stress_calc} for evaluation of the above expression for stress.

\section{Calculation of PNIPAm brush properties}
\label{stimuliEffect}
In this section, we present a numerical study of the effect of temperature, graft density and chain length on PNIPAm brush properties. The effect of temperature is taken into account in SST through Flory-Huggins parameter, $\chi$. For a classical polymer, $\chi=\chi(T)=\frac{\Theta}{2T}$, where $\Theta$ is $\theta$-temperature of the polymer solution, defined as the temperature at which excluded volume vanishes\cite{de79}. Notice that $\chi$ for a classical polymer depends only on temperature and thus is incapable of describing solution properties of an LCST polymer like PNIPAm. So, Flory-Huggins solution theory is generalized such that $\chi=\chi(T,\phi)$ which is assumed to have the following form: $\chi=\sum_{i=0}^n C_i(T)(\phi)^i$, where $C_i$ are functions of $T$, and $n$ is an integer\cite{huggins64,karlstrom1985,matsuyama1990,bekiranov1997,baulin2002}. The dependence of $\chi$ on $\phi$ may lead to a more complicated stimuli response of a brush such as  vertical phase separation into a polymer rich and a solvent rich phases within the brush\cite{baulin2003self,halperin2011}.

\begin{figure}[!h]
	\center
	\includegraphics[width=5.7cm]{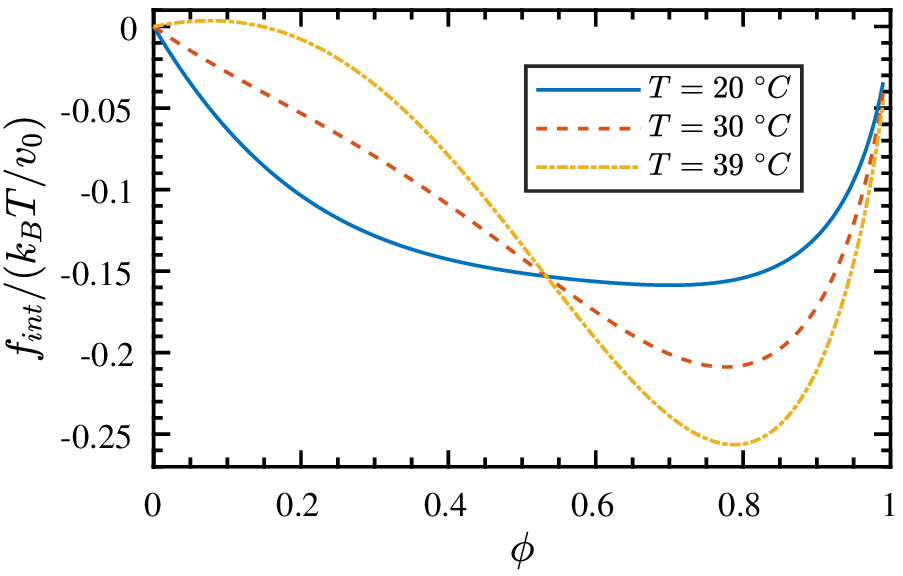}
	\hspace{-6mm}
	\includegraphics[width=5.7cm]{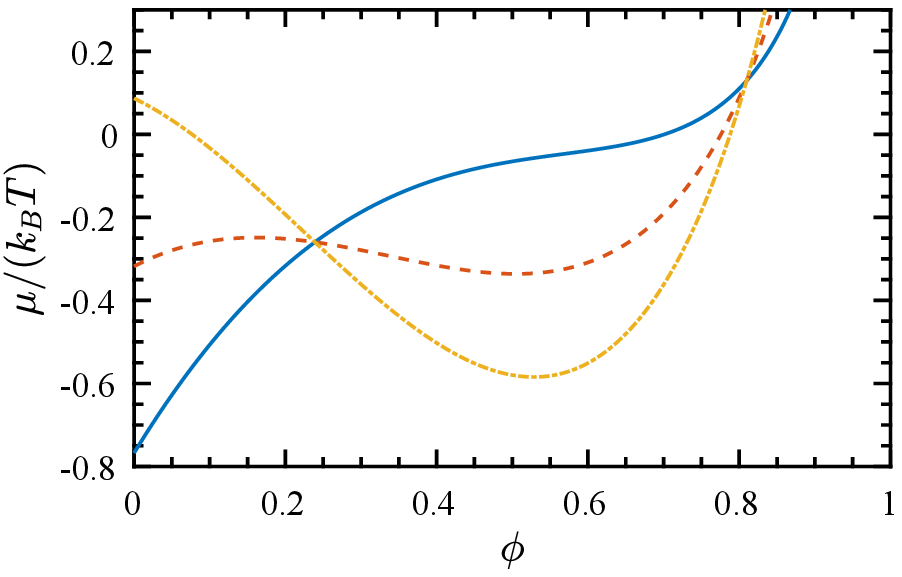}
	\hspace{-6mm}
	\includegraphics[width=5.7cm]{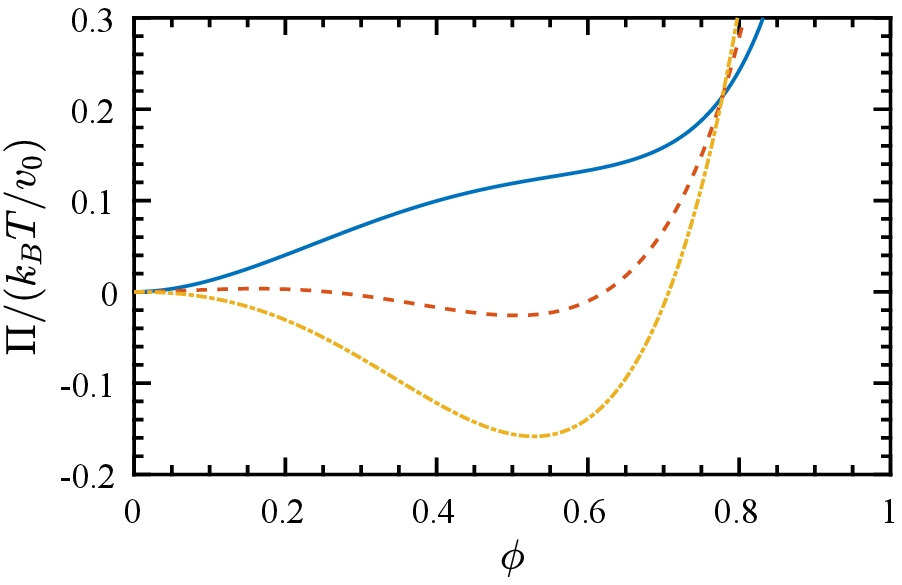}
	\caption{The variation of the interaction free energy density, the chemical potential, and the osmotic pressure in a PNIPAm-water solution with polymer volume fraction at different temperatures. It can be observed that the free energy curve at $20~^{\circ}C$ is convex and phase separation is will happen at this temperature. A possible phase segregation can be inferred from the negative slope in a part of the chemical potential curve corresponding to $30~^{\circ}C$. Observe large concavity in the free energy curve at $39~^{\circ}C$. They lead to globule formation.}
	\label{Fint_PNIPAm}
\end{figure}

Decades of research on the phase diagram of PNIPAm is yet to yield a definitive $\chi$ for PNIPAm \cite{halperin2015poly}. Here, we use $\chi$ of PNIPAm-water solution reported by Afroze \emph{et al.}\cite{afroze2000} as given below:
\begin{align}
\chi=\chi(T,\phi)&=C_0+ C_1\phi+C_2\phi^2, \quad \rm{where,} \nonumber\\
C_0&=-12.947 + 0.044959T,  \nonumber\\
C_1&=17.920-0.056944T, \nonumber\\
C_2&=14.814-0.051419T,
\label{eqch4p2}
\end{align}
and $T$ is in Kelvin. The above choice is guided by the fact that the vertical phase separation in a PNIPAm brush predicted by this form of $\chi$ has been observed experimentally\cite{yim2005,laloyaux2009,varma2016}. Interaction free energy density is calculated using Eq. \eqref{eq14}, from which chemical potential ($\mu(\phi)$) and osmotic pressure ($\Pi(\phi)$) are then obtained.
\begin{align}
\mu(\phi)&=v_0\frac{\partial f_{int}}{\partial \phi}=\left[-1-\ln(1-\phi)+(1-2\phi)\chi+\phi(1-\phi)\frac{\partial\chi}{\partial \phi}\right]k_BT, \label{eqch4p3a} \\
\Pi(\phi)&=\phi\frac{\partial f_{int}}{\partial\phi}-f_{int}=-\left[\phi+\ln(1-\phi)+\phi^2\chi-\phi^2(1-\phi)\frac{\partial\chi}{\partial \phi}\right]\frac{k_BT}{v_0},
\label{eqch4p3b}
\end{align}
\fref{Fint_PNIPAm} shows the variation of interaction free energy density,  chemical potential and osmotic pressure with volume fraction at different temperatures. When the interaction free energy \emph{vs} volume fraction curve at a given temperature $T$ has nonconvex sections (equivalently, the chemical potential curve has sections with negative slope), a homogeneous polymer solution with the volume fraction in that section becomes thermodynamically unstable, and the free energy is minimized by phase segregation\cite{teraoka2002polymer}. The polymer solution may segregate into a polymer rich phase and a solvent rich phase, or it can form globules separated from pure solvent. In the first case, the free energy is minimized when the polymer volume fractions in the two phases simultaneously satisfy the conditions that the chemical potential and the osmotic pressure in the two phases be equal: $\mu(\phi^+(T))=\mu(\phi^-(T))$ and $\Pi(\phi^+(T))=\Pi(\phi^-(T))$, where $\phi^+(T)$ and $\phi^-(T)$ are the volume fractions in the polymer rich and the solvent rich phases at temperature $T$, respectively. Also, in this case, the occurrence of $\phi^-(T)<\phi<\phi^+(T)$ is prohibited in a thermodynamic equilibrium. In the second case, the free energy is minimized when the polymer volume fraction in a globule satisfies the criterion that the osmotic pressure in the globule be zero ($\Pi(\phi^+(T)=0$, where $\phi^+(T)$ is the polymer volume fraction in the globule at temperature $T$). In this case, $\phi<\phi^+(T)$ cannot occur in thermodynamic equilibrium.

Observe in \fref{Fint_PNIPAm} that at temperatures below the critical temperature, $T_{cr}=26.36~^{\circ}C$, the solvent is a good solvent ($f_{int}$ is always convex for all $\phi$). Above $30.5~^{\circ}C$, the polymer chains form globules. Between these two temperatures, a segregation into polymer rich and solvent rich phases occurs. This is visible from the negative slope in parts of the chemical potential curve for $T=30~^{\circ}C$.

Calculation of the stress in a brush using \eqref{eq18} and \eqref{eq19f} requires us to know the polymer volume fraction and the end density which is used to calculate the free energy density, and the strain ratio $\frac{\partial\epsilon_{\textsc{zz}}}{\partial\epsilon_{\textsc{xx}}}$ in the brush. In the following sections, we obtain these properties for a low, a moderate and a high graft density brush at different temperatures. We use them to finally calculate stress and resultant properties. Note that no definitive values for the structural parameters of a PNIPAm polymer chain are available in the literature. In the calculations in the sections below, we assume $r_k=1$.

\subsection{Polymer volume fraction}
\label{volume_fraction}
The polymer volume fraction within a brush can be calculated using Eq. \eqref{eq1}. It is convenient to use the following relation\cite{baulin2003self,halperin2011} to obtain the volume fraction:
\begin{equation}
\bar{\mu}(\bar{\phi}(0))-\bar{\mu}(\bar{\phi}(\bar{\textsc{z}}))=\bar{V}(\bar{\textsc{z}}),
\label{eqch4p4}
\end{equation}
which is obtained by taking the difference of Eq. \eqref{eq1} with the equation obtained by substituting $\bar{\textsc{z}}=0$ in Eq. \eqref{eq1}. To obtain $\bar{\phi}(\bar{\textsc{z}})$ at a given temperature $T$, we first choose $\bar{\phi}(0)$ and then use Eq.\eqref{eqch4p4}. Additionally, if $26.36~^{\circ}C<T<30.5~^{\circ}C$, $\phi^-(T)<\phi<\phi^+(T)$ cannot occur. If $\bar{\phi}(0)<\phi^-(T)$, then volume fraction profile will still be smooth. However, if $\bar{\phi}(0)>\phi^+(T)$, vertical phase separation within the brush occurs. Also, for $T>30.5~^{\circ}C$, $\phi<\phi^+(T)$ is not allowed. Hence, a jump in the volume fraction profile at the brush free end occurs. Graft density is obtained from $\bar{\phi}(\bar{\textsc{z}})$ by using the relation
\begin{equation}
\rho_g=\frac{a}{v_0}\int_0^{\bar{H}}\bar{\phi}(\bar{\textsc{z}})d\bar{\textsc{z}}.
\label{eqch4p4b}
\end{equation}
We nondimensionalize graft density by defining $\bar{\rho}_g=\rho_g v_0/a$.
\begin{figure}[!h]
	\center
	\includegraphics[width=5.7cm]{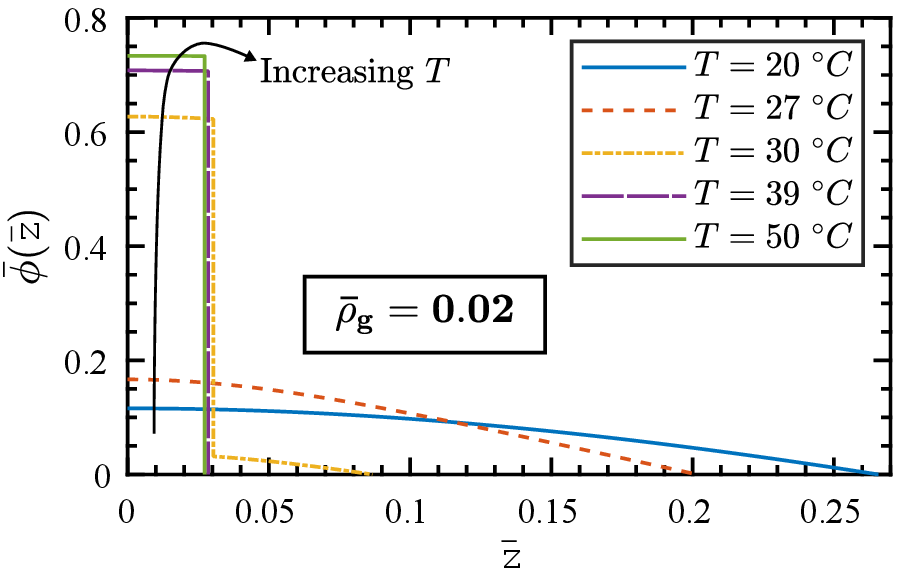}
	\hspace{-6mm}
	\includegraphics[width=5.7cm]{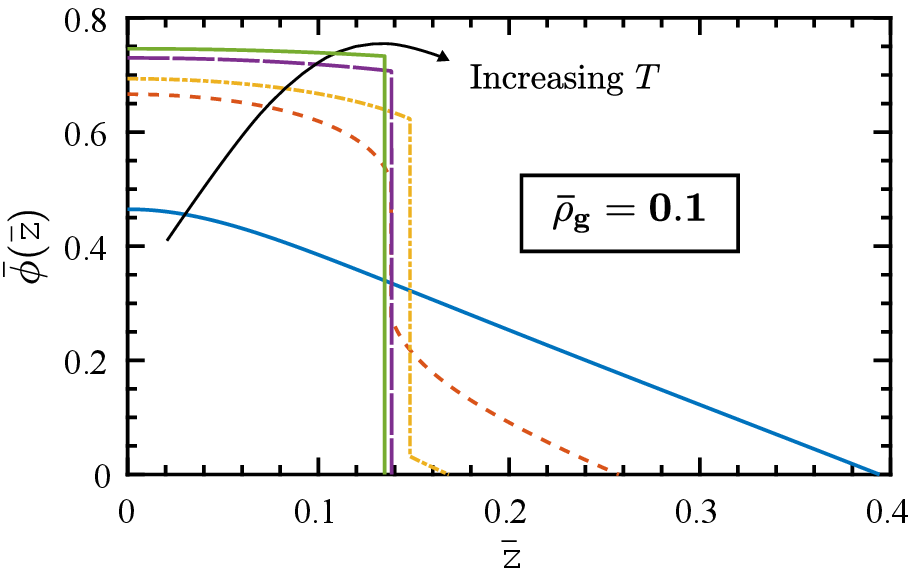} 
	\hspace{-6mm}
	\includegraphics[width=5.7cm]{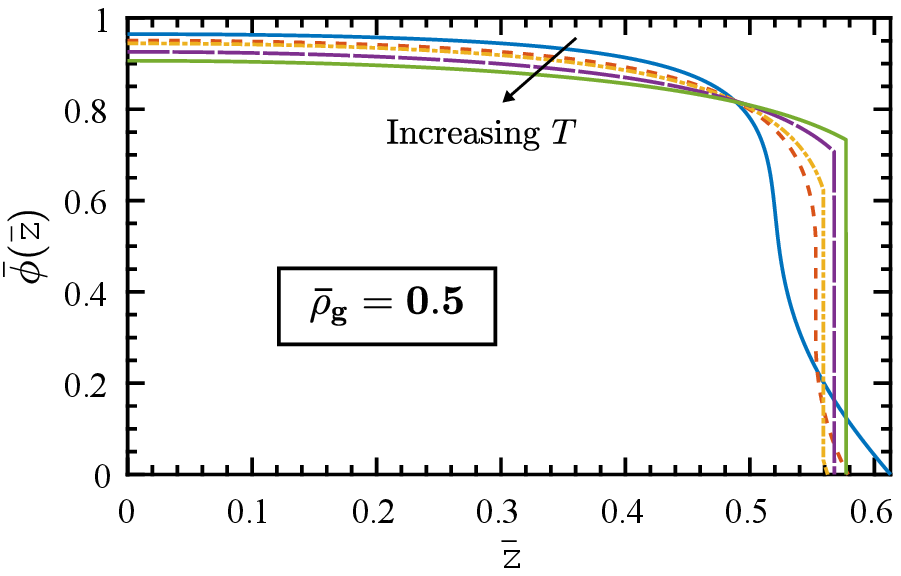}
	\caption{The variation of the polymer volume fraction profile with temperature for three graft densities. Notice the reversal in trend in the plot for the highest graft density plotted.}
	\label{numden_PNIPAm}
\end{figure}
Below the critical temperature ($ T_{cr}=26.36~^{\circ}C$), the volume fraction profile is continuous and reaches zero at the brush free end (see \fref{numden_PNIPAm}). Also, the profile is close to parabolic, like in a brush of gaussian chains, for the lowest graft density brush shown in \fref{numden_PNIPAm}. The profile changes with increasing temperature and becomes step like at temperatures above $30.5~^{\circ}C$.  In between these limiting temperatures, a profile with a vertical phase separation may occur. Notice the drastic change in the profile for the lower graft densities in \fref{numden_PNIPAm}. Remarkably, the effect of temperature on the profile for the highest graft density is minimal and the profile is step like away from the brush free end even in good solvent conditions. See Appendix~\ref{H_SR} for a discussion on the effect of temperature on the brush height and the swelling ratio of different graft density brushes.

\begin{figure}[!h]
	\center
	\includegraphics[width=8cm]{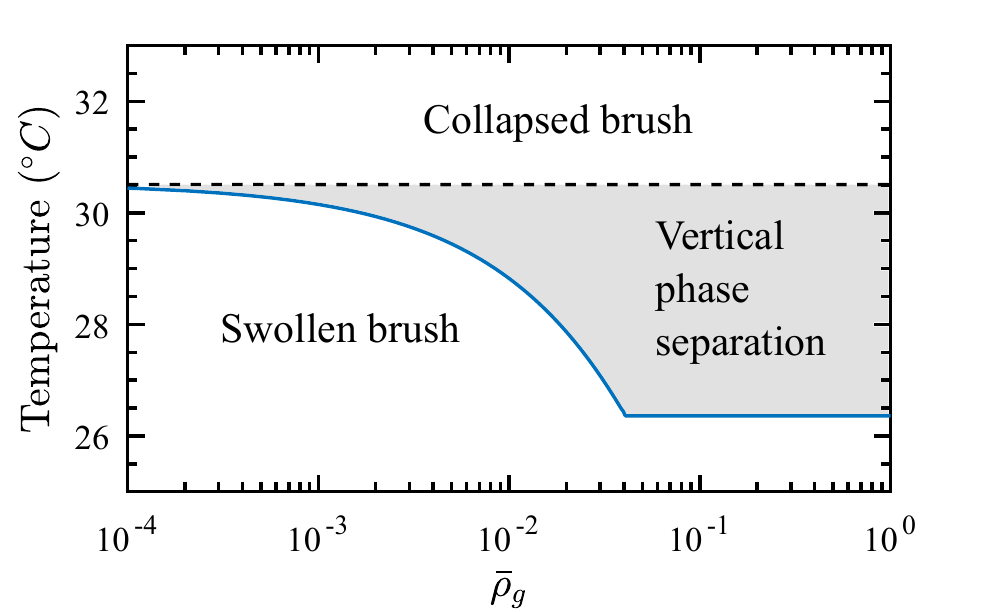}
	\caption{Phase boundaries for PNIPAm brushes. Vertical phase separation within a brush with  $\bar{\rho}_g>0.04$ starts right above the critical temperature. For $\bar{\rho}_g<0.04$, it depends on graft density with an asymptote at $30.5~^{\circ}C$. Above $30.5~^{\circ}C$, all the brushes collapse.}
	\label{phaseBoundary}
\end{figure}

Vertical phase separation within a brush occurs only if $\phi(0)>\phi^+(T)$ at $26.36~^{\circ}C<T<30.5~^{\circ}C$. Hence, although all the brushes show a vertical phase separation within the brush above a certain temperature, the onset of such a separation depends on $\phi(0)$ and consequently graft density as shown in \fref{phaseBoundary}. \fref{phaseBoundary} also suggests that a very low graft density brush would collapse from its swollen form with only a small change in temperature. Therefore, a vertical phase separation in them would be difficult to observe experimentally. Moreover, for the brushes with $\bar{\rho}_g>0.04$, $\bar{\phi}(0)>\phi^-(T)$ corresponding to any temperature $>T_{cr}$. Hence, for these brushes, boundary defining onset of phase separation is at $T_{cr}$.

\subsection{End density}
\label{end_density}
\begin{figure}[!h]
	\center
	\includegraphics[width=5.7cm]{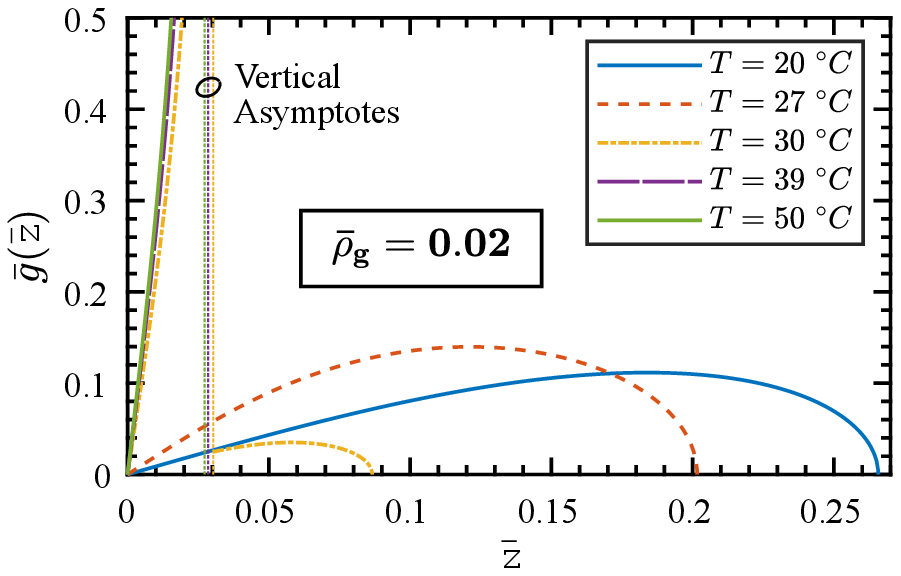}
	\hspace{-6mm}
	\includegraphics[width=5.7cm]{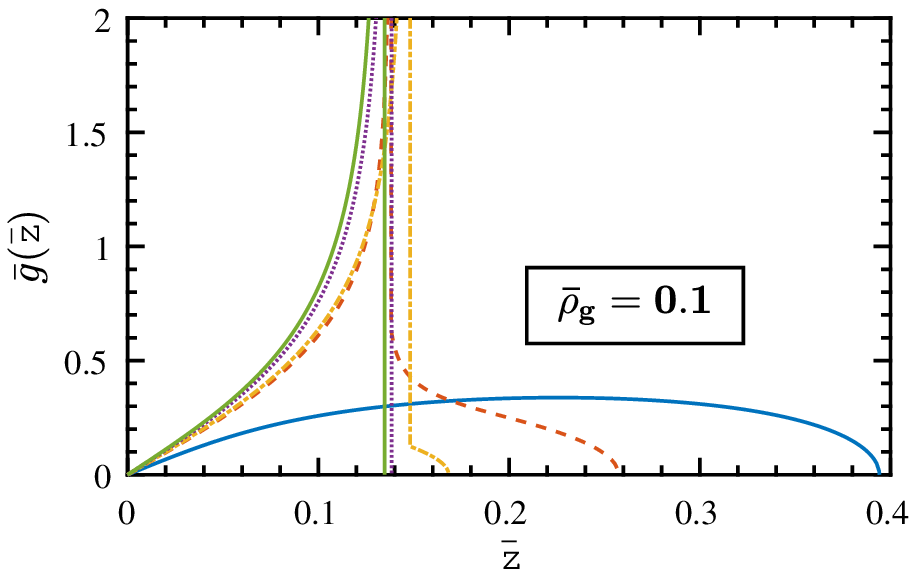}
	\hspace{-6mm}
	\includegraphics[width=5.7cm]{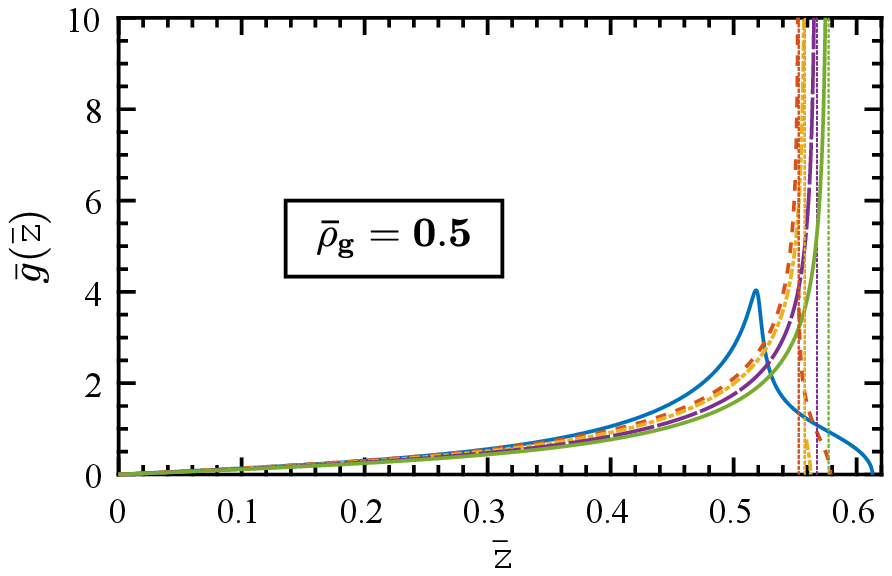}
	\caption{The variation of end density profile with temperature. Vertical asymptotes indicate diverging end density at the location where a jump in the polymer volume fraction profile is observed.}
	\label{endden_rhog_Tvary_PNIPAm}
\end{figure}
The end density profile at different temperatures are shown in \fref{endden_rhog_Tvary_PNIPAm} for three graft densities. At the temperatures below the critical temperature for PNIPAm, for low graft density brushes, the profile is a smooth curve like in a Gaussian chain brush. However, the profile has a sharp peak for the highest graft density brush even below the critical temperature. The sharp peak is due to a sharp change in polymer volume fraction profile (high $\frac{d\bar{\phi}(\bar{\textsc{z}})}{d\bar{\textsc{z}}}$). At the location of discontinuity within the volume fraction profile, end density diverges and a vertical asymptote is observed. As temperature increases further, the end density at $\bar{\textsc{z}}>\bar{H}_t$ decreases and becomes zero at temperatures above $30.5~^{\circ}C$.

\subsection{Osmotic pressure}
\label{osmotic_pressure}
\begin{figure}[!h]
	\center
	\includegraphics[width=5.7cm]{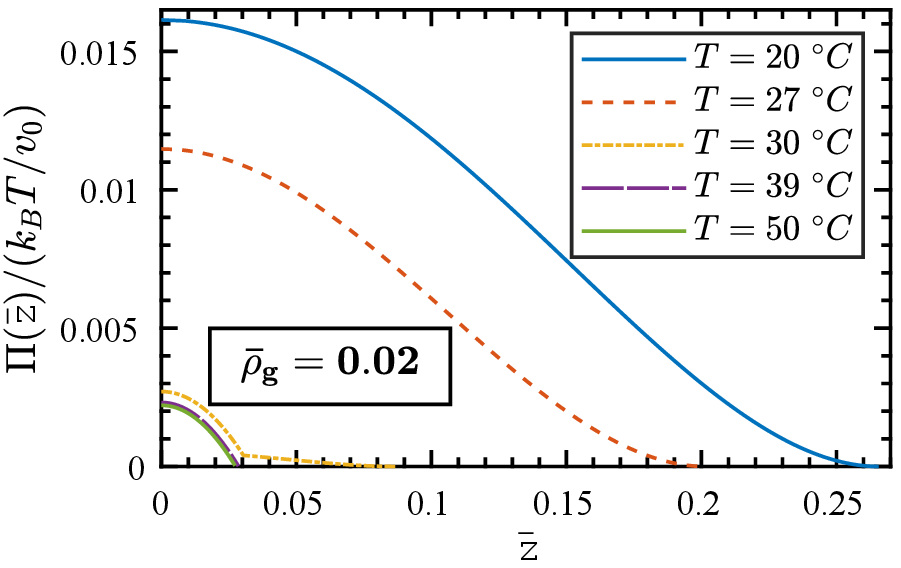}
	\hspace{-6mm}
	\includegraphics[width=5.7cm]{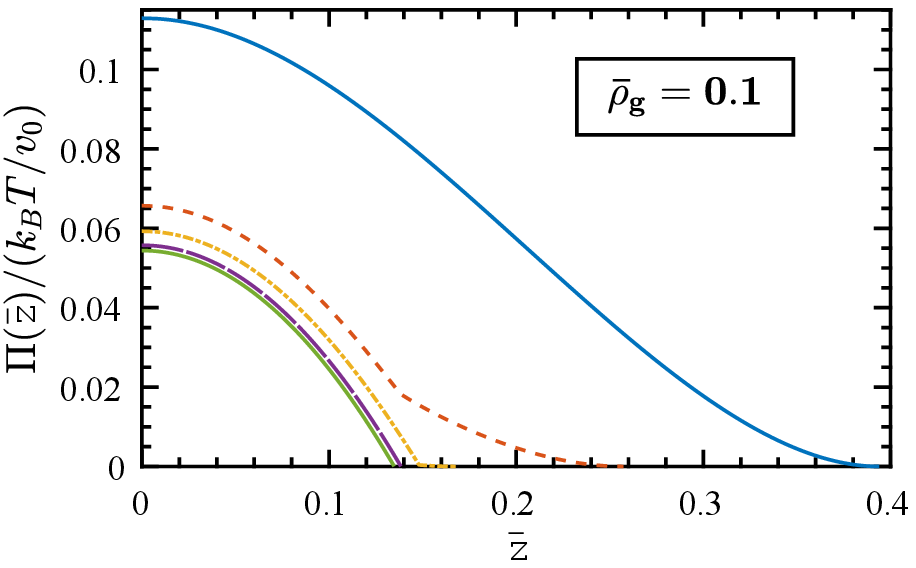}
	\hspace{-6mm}
	\includegraphics[width=5.7cm]{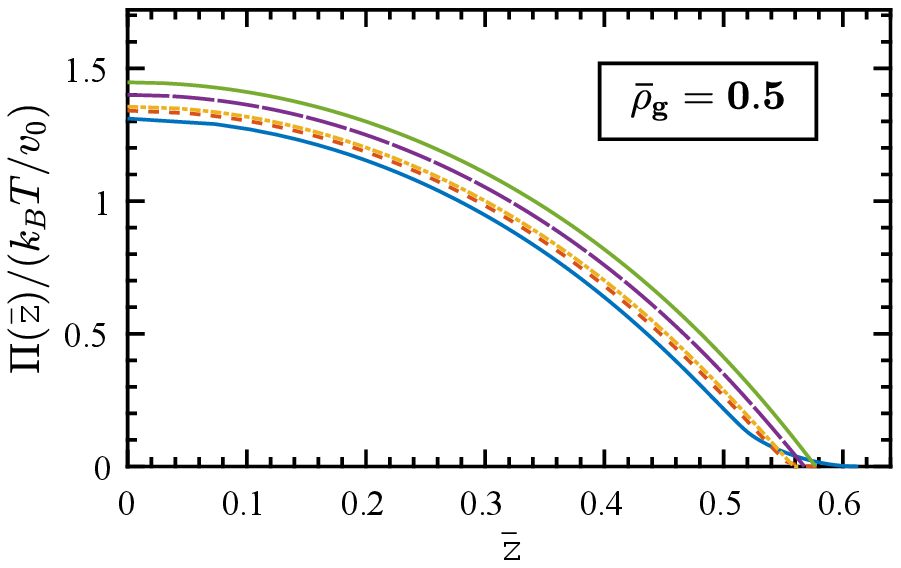}
	\caption{The variation of the osmotic pressure within a brush at different temperatures for brushes of low, moderate and high graft densities.}
	\label{P_os_PNIPAm}
\end{figure}
The osmotic pressure within a brush decreases sharply with increasing temperature up to $33~^{\circ}C$ and less so above this temperature, for low graft density brushes (see \fref{P_os_PNIPAm}). For very high graft density brushes, unlike low to moderate density brushes, the osmotic pressure shows a mild increase with an increase in temperature as seen for $\bar{\rho}_g=0.5$ in \fref{P_os_PNIPAm}. As expected, the osmotic pressure decreases monotonically with the distance from the grafting surface and is continuous, though it may have a corner for a small range of temperature near the critical temperature.

\subsection{Strain ratio $\frac{\partial\epsilon_{\textsc{zz}}}{\partial\epsilon_{\textsc{xx}}}$}
\label{strain}
\begin{figure}[!h]
	\center
	\includegraphics[width=5.7cm]{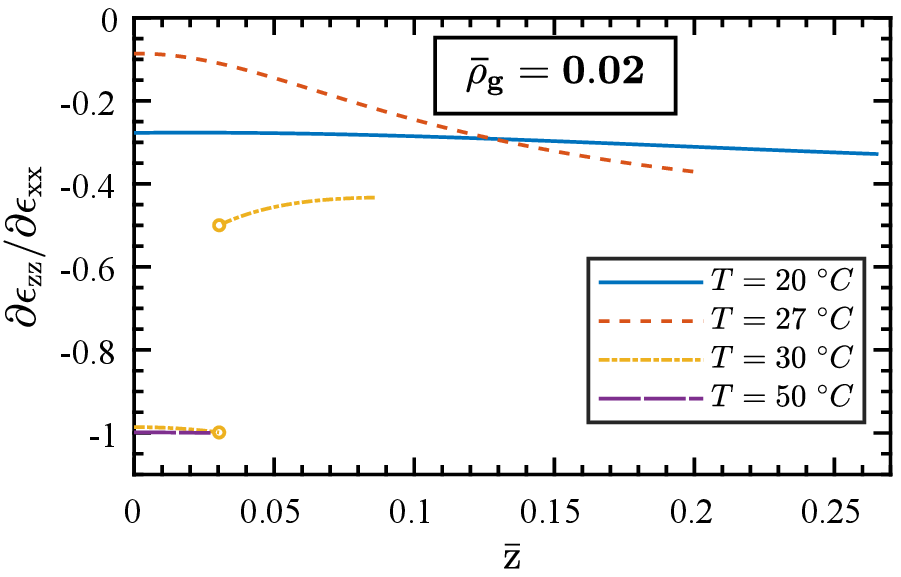}
	\hspace{-6mm}
	\includegraphics[width=5.7cm]{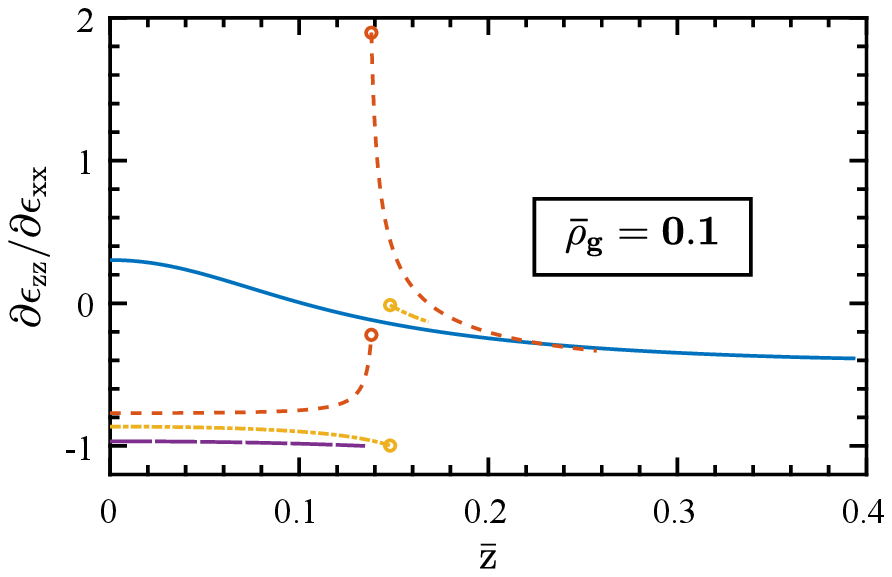}
	\hspace{-6mm}
	\includegraphics[width=5.7cm]{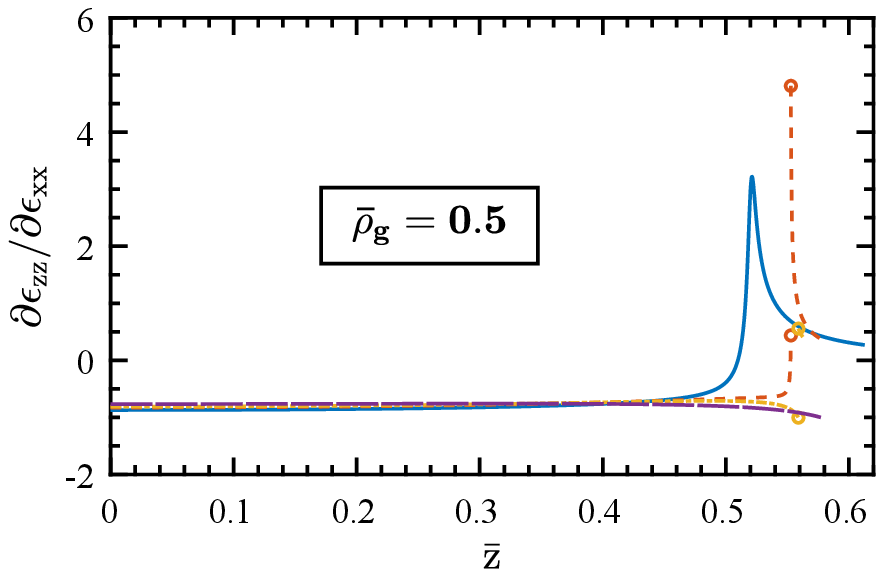}
	\caption{The strain ratio $\frac{\partial\epsilon_{\textsc{zz}}}{\partial\epsilon_{\textsc{xx}}}$ in a layer is plotted \emph{vs} the distance from the grafting surface for three graft densities. The discontinuity in volume fraction profile leads to a discontinuity in $\frac{\partial\epsilon_{\textsc{zz}}}{\partial\epsilon_{\textsc{xx}}}$ curve, which is not defined at $\bar{H}_t$, as denoted by circles. Note that the curves at $39~^{\circ}C$ are very close to the curves at $50~^{\circ}C$, and are not shown.}
	\label{dmuByde_rhog_Tvary_PNIPAm}
\end{figure}

Calculation of the stress within a brush requires us to evaluate the strain $\delta\epsilon_{\textsc{zz}}$ in the brush as an infinitesimal external strain $\delta\epsilon_{\textsc{xx}}$ is applied. \fref{dmuByde_rhog_Tvary_PNIPAm} shows the profile of $\frac{\partial\epsilon_{\textsc{zz}}}{\partial\epsilon_{\textsc{xx}}}$ at different temperatures for three graft densities. The discontinuity in the brush volume fraction profile leads to a discontinuity in these curves as well. A negative value of $\frac{\partial\epsilon_{\textsc{zz}}}{\partial\epsilon_{\textsc{xx}}}$ corresponds to shrinking of a layer in the $\textsc{z}$-direction as the brush is stretched in the $\textsc{x}$-direction. For a low density brush in a good solvent, $\frac{\partial\epsilon_{\textsc{zz}}}{\partial\epsilon_{\textsc{xx}}}$ is close to $-1/3$, the value predicted by the Gaussian chain SST\cite{manav2018}. For moderate density brushes, this ratio becomes positive close to the grafting surface, signifying an \emph{expansion} of a thin layer in the $\textsc{z}$-direction as the brush is stretched in the $\textsc{x}$-direction. For high density brushes, the curves are only weakly affected by a change in temperature for the most of the brush, except near the brush free end, where a layer expands instead of shrinking below $30.5~^{\circ}C$.

\subsection{Stress}
\label{stress_PNIPAm}
\begin{figure}[!h]
	\center
	\includegraphics[width=5.7cm]{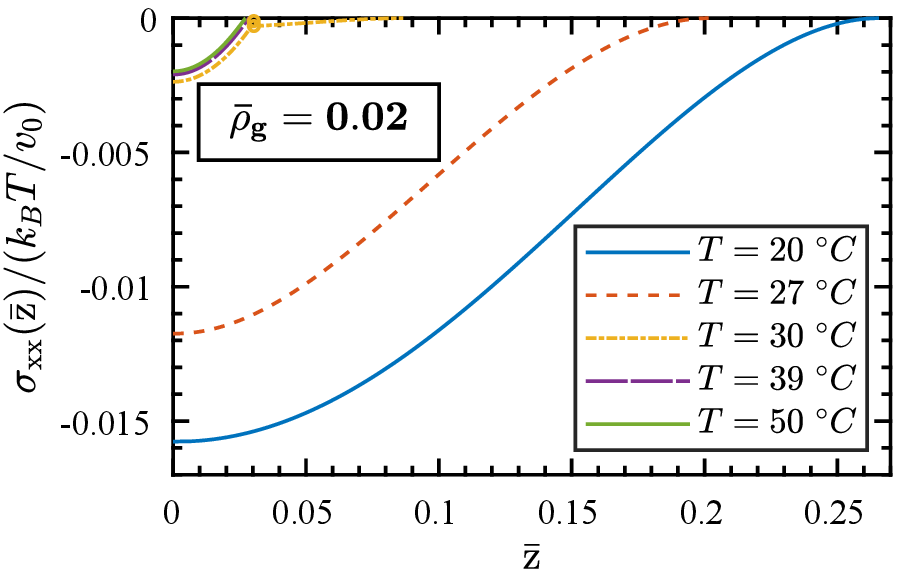}
	\hspace{-6mm}
	\includegraphics[width=5.7cm]{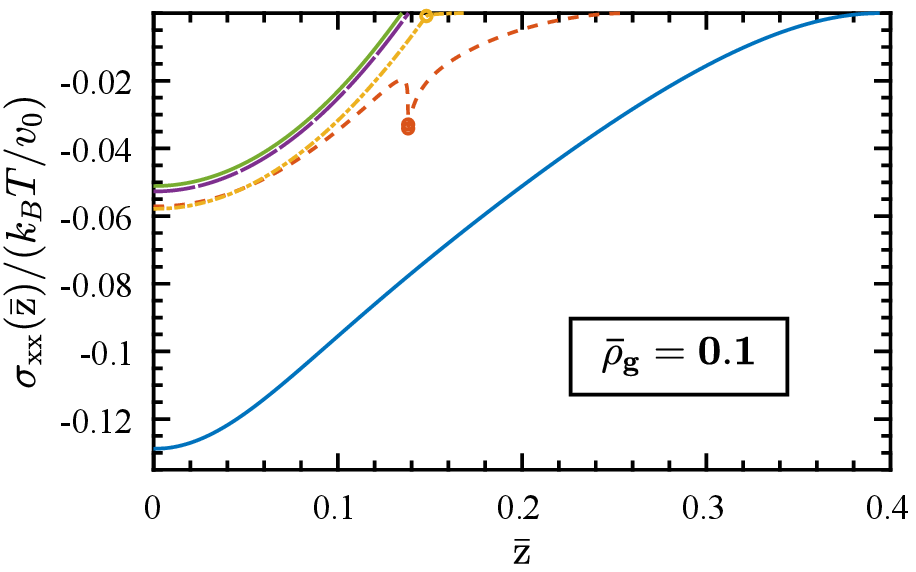}
	\hspace{-6mm}
	\includegraphics[width=5.7cm]{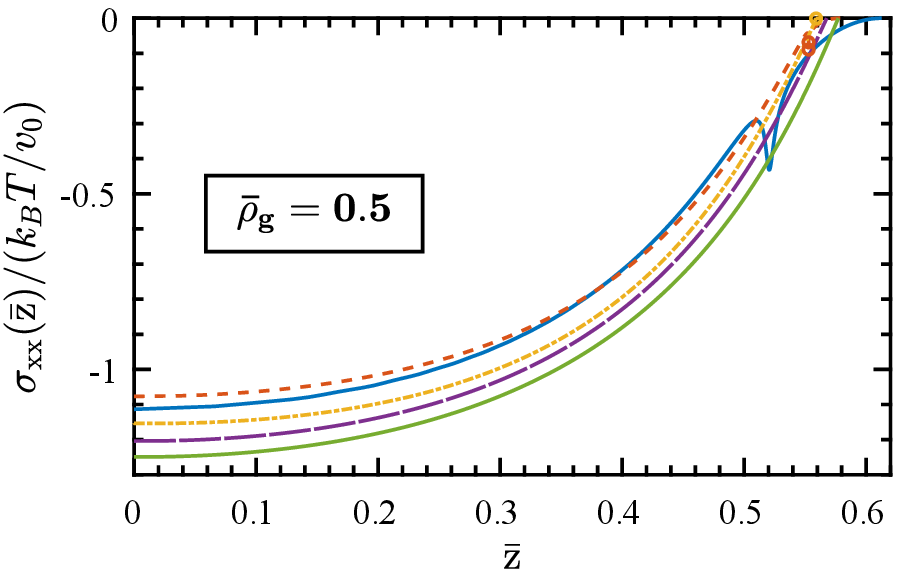}
	\caption{The variation of the stress within the brushes of three graft densities at different temperatures. Circles denote that the stress is not defined at the point.}
	\label{Sxx_z_PNIPAm}
\end{figure}
Now we have all the brush properties needed to evaluate the stress in a brush. The stress variation for three graft densities is shown in \fref{Sxx_z_PNIPAm}. Like other brush properties, the stress within a PNIPAm brush can be tuned by changing temperature. However, it is always \emph{compressive}, even at temperatures $>30.5~^{\circ}C$ when the solvent becomes poor and the brush collapses. Also, unlike the volume fraction profile, which approaches step profile as temperature is raised above $30.5~^{\circ}C$, stress retains inhomogeneity at all temperatures. The stress profiles at temperatures above $30.5~^{\circ}C$ show a monotonic decrease in the magnitude of stress as the distance from the grafting surface increases. This however is not always true at a temperature below $30.5~^{\circ}C$.

The variation in stress is found to be approximately a quartic function of the distance from the grafting surface for the lowest graft density at temperatures below the critical temperature, as predicted by the SST with Gaussian chains\cite{manav2018}. The stress profile changes considerably with increasing temperature up to $33~^{\circ}C$ for low to moderate density brushes. For high density brushes however, the stress profile is minimally affected by a change in temperature in comparison to lower graft density brushes. In a brush with vertical phase separation inside the brush, the stress profile is discontinuous. At the interface, stress in undefined because the numerator in the expression for $\sigma_{\textsc{xx}}$ in \eqref{eq19f} is found to be finite in numerical calculation.

The stress in a brush causes the bending of a flexible substrate. The bending of a substrate much thicker than the brush is governed by the resultant stress due to the brush which is defined in the next section. For a substrate with thickness similar to the brush height, the resultant bending moment also affects the bending of the substrate. In the following sections, we look at the variation of the resultant stress and the resultant moment due to brushes of different graft densities with temperature. To demonstrate the bending of the substrates caused by the brushes, an example of the bending of two brush grafted substrates with thicknesses (a) similar to the maximum brush height, and (b) much greater than the maximum brush height is also discussed.

\subsubsection{Resultant stress}
Resultant stress, defined as
\begin{equation}
\tau_s=Na\int_0^{\bar{H}}\sigma(\bar{\textsc{z}})d\bar{\textsc{z}},
\label{res_Sxx}
\end{equation} 
which also equals $\frac{1}{A}\frac{\partial}{\partial\epsilon_{\textsc{xx}}}(A(F_{int}+F_{el}))=-\bar{\rho}_g^2\frac{\partial}{\partial\bar{\rho}_g}\left(\frac{F_{int}+F_{el}}{\bar{\rho}_g}\right)$, where $A$ is substrate surface area, was also calculated. It is compressive and increases in magnitude nonlinearly with increasing graft density and linearly with increasing chain length. It also exhibits rapid decrease in magnitude up to $33~^{\circ}C$ and plateaus after that for $\bar{\rho}_g<0.3$. For $\bar{\rho}_g>0.3$, in contrast, $\tau_s$ shows a moderate increase in magnitude with increasing temperature, that is as the solvent becomes poorer. 
\begin{figure}[!h]
	\center
	\includegraphics[width=8cm]{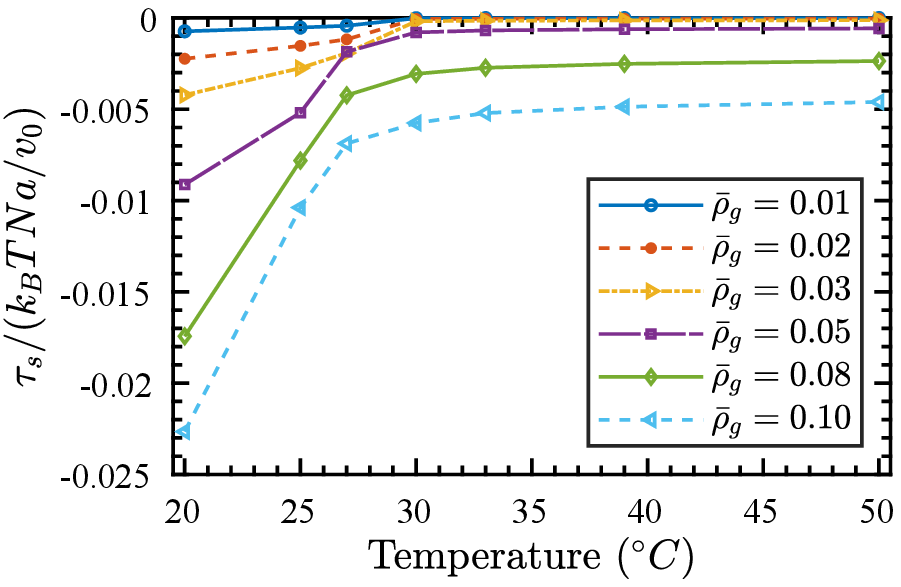}
	\hspace{-8mm}
	\includegraphics[width=8cm]{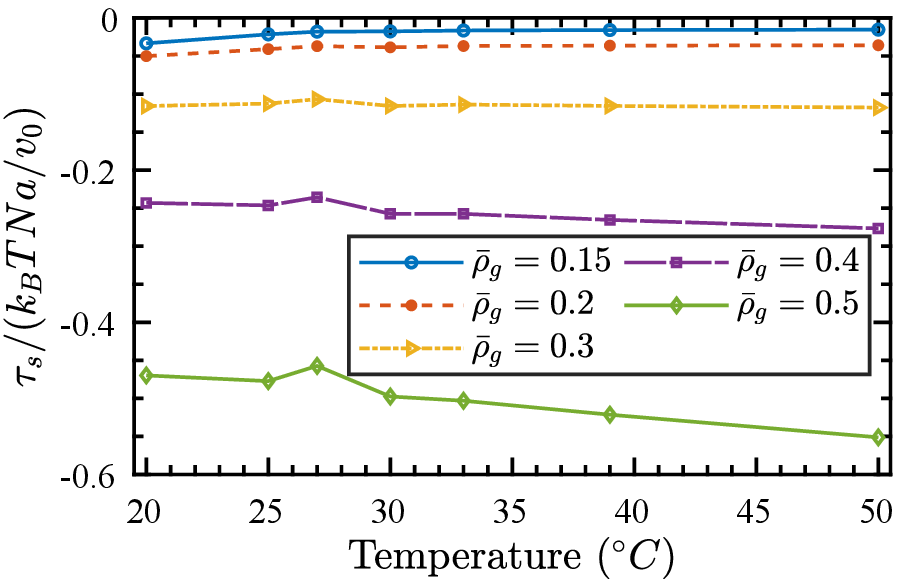}
	\caption{The variation of the resultant stress ($\tau_s$) with temperature at different graft densities. For lower graft densities ($\bar{\rho}_g<0.3$), $\tau_s$ decreases in magnitude with increasing temperature and reaches a plateau at $33~^{\circ} C$. For Higher graft densities, $\tau_s$ increases in magnitude with increasing temperature. Note that a nonsmooth change in the $\tau_s$ curves at $27~^{\circ} C$ results from difficulty in numerical calculation due to vertical phase separation within the brush.}
	\label{surfS_T_rhogvary_PNIPAm}
\end{figure}

\begin{figure}[!h]
	\center
	\includegraphics[width=8cm]{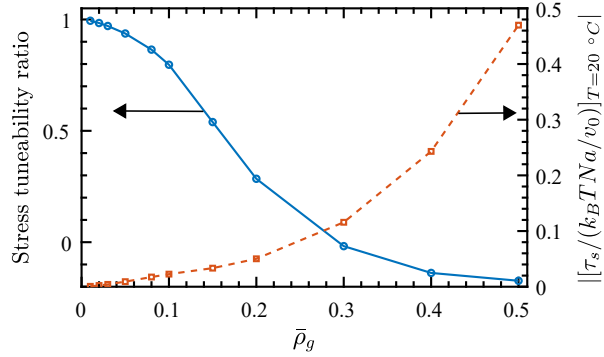}
	\caption{On the left y-axis, stress tuneability ratio, the ability to tune stress by changing temperature from $20^{\circ}C$ to $50^{\circ}C$, is plotted \emph{vs} graft density. On the right y-axis, the magnitude of the resultant stress at $20~^{\circ}C$ is plotted.}
	\label{stressTuneability_PNIPAm}
\end{figure}

Furthermore, we plot stress tuneability ratio, defined as $1-[\tau_s]_{50^{\circ}C}/[\tau_s]_{20^{\circ}C}$, where $[\tau_s]_{T}$ is the resultant stress at temperature $T$, as a function of graft density (see \fref{stressTuneability_PNIPAm}). It decreases with increase in graft density and the sharpest decline is observed near $\bar{\rho}_g=0.15$. This is in contrast with resultant stress which increases in magnitude with increasing graft density. So, a trade off exists between the maximum compressive resultant stress and the ability to modify it by varying temperature. Also, for $\bar{\rho}_g\ge0.3$, the ratio is negative.

\subsubsection{Resultant bending moment}
\begin{figure}[!h]
	\center
	\includegraphics[width=8cm]{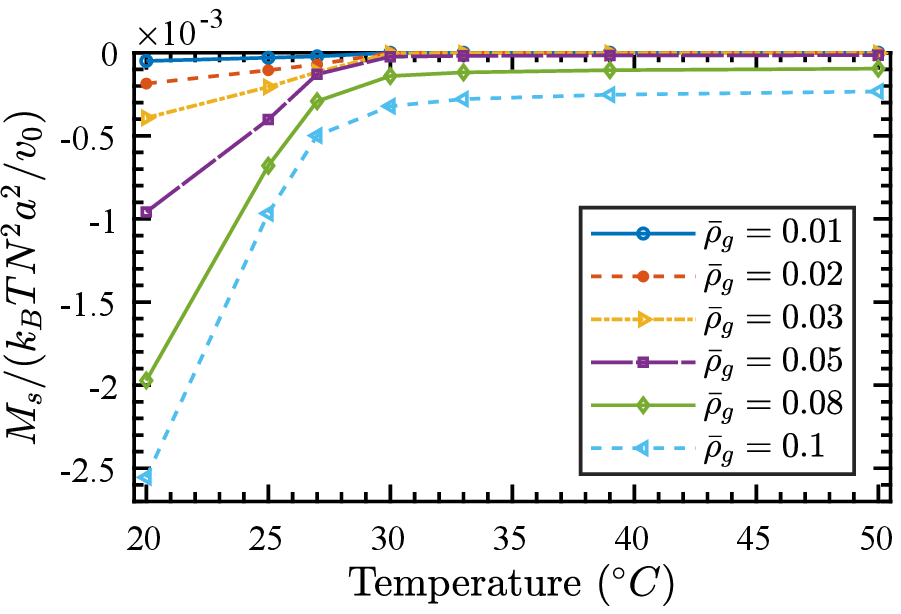}
	\hspace{-8mm}
	\includegraphics[width=8cm]{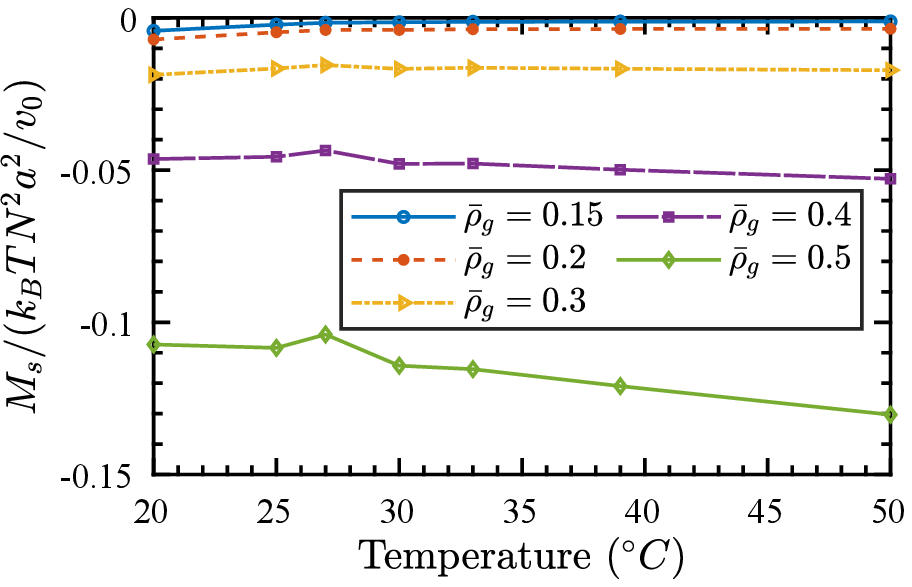}
	\caption{The variation of the resultant moment with temperature at different graft densities.}
	\label{Ms_T_rhogvary_PNIPAm}
\end{figure}

Resultant bending moment is defined as
\begin{equation}
M_s=(Na)^2\int_0^{\bar{H}}\sigma(\bar{\textsc{z}})\bar{\textsc{z}}d\bar{\textsc{z}}.
\label{res_Mxx}
\end{equation}
\fref{Ms_T_rhogvary_PNIPAm} plots the variation of the resultant moment with graft density at different temperatures. Resultant moment also increases in magnitude with increasing graft density. It increases in magnitude with square of chain length. Its temperature dependence is qualitatively similar to that of resultant stress. For $\bar{\rho}_g<0.3$, it decreases in magnitude with increasing temperature. Like $\tau_s$, the pattern is reversed for $\bar{\rho}_g>0.3$.

\subsection{Tuning of substrate bending}
\begin{figure}[!h]
	\center
	\includegraphics[width=8cm]{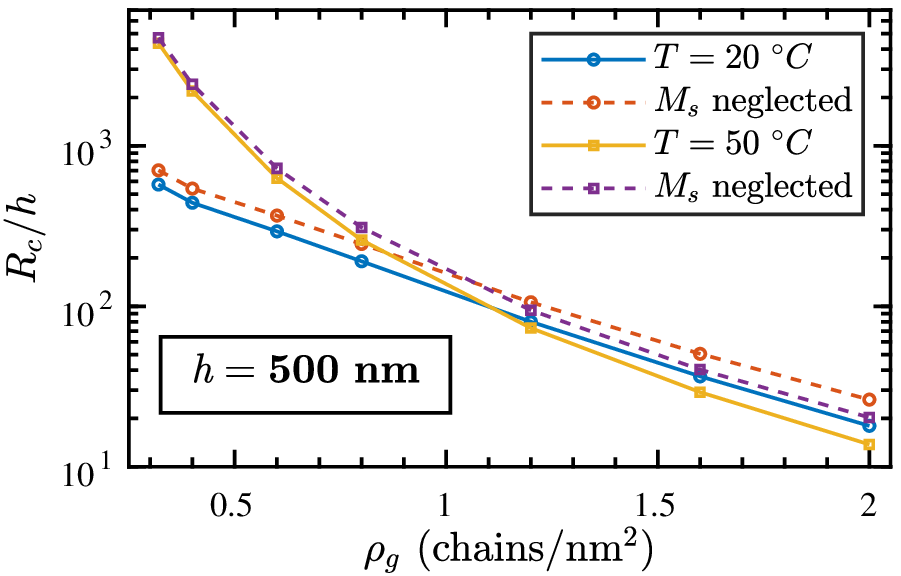}
	\hspace{-8mm}
	\includegraphics[width=8cm]{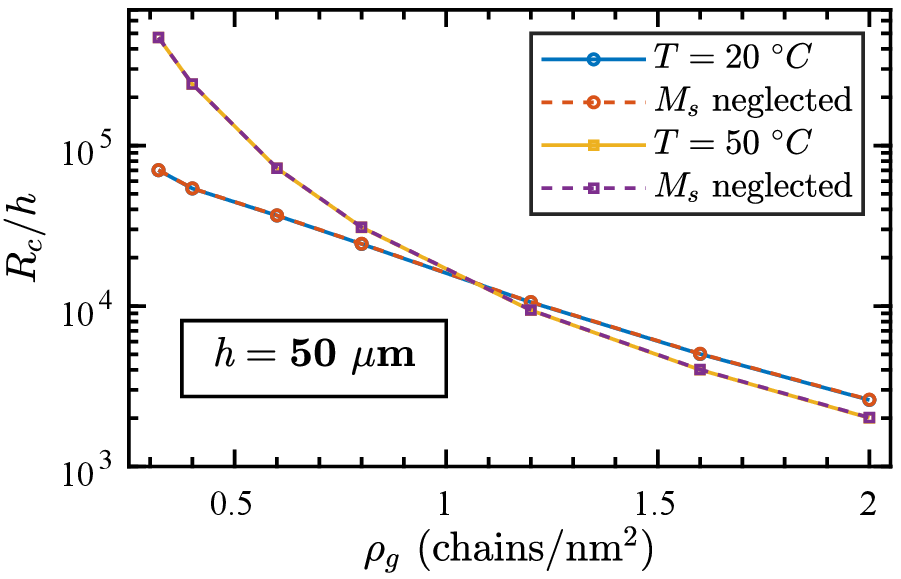}
	\caption{The radius of curvature (normalized by substrate thickness) of a bent substrate for two different thicknesses. Dashed curves are obtained by substituting $M_s=0$ in Eq. \eqref{Sxx_R_rel}.}
	\label{Rc_rhog_Tvary_PNIPAm}
\end{figure}

In this section, we discuss an example of bending of a substrate due to polymer brush, and the change in the configuration of the substrate on changing temperature. In the first step, we need values of monomer size, Kunh length, and monomer volume for PNIPAm. Since the experimental measurements have not yielded a consistent value\cite{halperin2011}, we choose $a=0.5$ nm, $r_k=1$, and $v=a^3=0.125$ nm$^3$. Also, we assume $N=1000$. In one effective monomer there will be approximately two repeat units of PNIPAm, each of molecular weight $113.16$~g/mol. Hence molecular weight of a chain is $\sim 2\times10^5$~g/mol, which is within the range of values seen in experiments. Then, brush height even at the maximum graft density studied is $H<0.65Na=325$~nm. We assume that the Young's modulus of the substrate is $E=2$~GPa and Poisson's ratio is $\nu=0.4$. Also, we choose two different substrates of thicknesses $h=500~$nm and $h=50~\mu$m. Then, using the following equation derived in Manav \emph{et al.}\cite{manav2018}, we obtain the radius of curvature, $R_c$.
\begin{equation}
\frac{h}{R_c}=-\frac{6\tau_s h+12M_s}{\bar{E}h^2-\frac{\nu}{1-\nu}\tau_s h},\quad \bar{E}=\frac{E}{1-\nu^2}, \quad i=1,~2.
\label{Sxx_R_rel}
\end{equation}
Note that in the above equation, unlike Manav \emph{et al.}\cite{manav2018}, $M_s$ has not been ignored. However, we ignore Young's modulus and other material constants for polymer brush resulting in $\tau_s$ and $M_s$ being independent of the state of bending.  The effective Young's modulus of a brush layer is of the same order of magnitude as the resultant stress. Since the strain in bending is small, the Young's modulus has negligible effect on bending. These assumptions are valid only for the case of small strain in the brush-substrate system, which is also required for the validity of the above equation. Notice that the above equation is distinct from the classical stoney equation ($h/R_c=-6\tau_s/(\bar{E}h)$). Furthermore, a brush can be assumed planar only if $\frac{H}{R_c}<<1$. \fref{Rc_rhog_Tvary_PNIPAm} shows the variation of the radius of curvature (nondimensionalized by dividing by substrate thickness) with graft density for the two substrates. Notice that at $\rho_g\approx 1~\rm{chains}/nm^2$, substrate curvature is unaffected by temperature change. Below this graft density, the substrate flattens on increasing temperature from $20~^{\circ}C$ to $50~^{\circ}C$. Also, the bending moment due to brush affects the bending of a thin substrate with thickness comparable to the brush. For a thicker substrate however, it can be ignored.

Notice that the above plots show results only for graft densities $>0.3$~chains/nm$^2$. This lower limit on graft density has been set by validity of strong stretching assumption, which holds only when stretching parameter~\cite{netz98,manav2019} $\beta_s=\frac{3}{2}\frac{H^2}{N r_k a^2}=\frac{3}{2}\frac{N}{r_k}\bar{H}^2$ is larger than a threshold value (taken to be 15 here). For low graft density brushes in a poor solvent, stretching parameter for $N=1000$ is not sufficiently large to ensure validity of this assumption. Utz~\emph{et al.}\cite{utz08} have argued that at low graft densities when chains in a brush form globules in a poor solvent, lateral stress also originates from surface tension at the globule-solvent and globule-substrate interfaces. This has not been studied in this work.

\section{Discussion}
\label{discussion}
Numerical results for PNIPAm brush properties suggest that the properties are strongly affected by temperature at low graft densities. Increasing graft density leads to weakening of the effect of temperature. For very high graft density brushes, the trend in the effect of temperature on brush properties is reversed in comparison to lower graft density brushes. This is a consequence of an interplay between the interaction free energy density for PNIPAm, which has a minima at a very high polymer volume fraction, and the stiffening of chains at large extensions due to their finite extensibility.

The polymer volume fraction profile of a low to moderate graft density brush is considerably altered by a change in temperature up to $33~^{\circ}C$ above which the change is small. This results in a large swelling ratio at $20~^{\circ}C$ (see \fref{swellingRatio_T_rhogvary_PNIPAm} in Appendix~\ref{H_SR}). However, the profile for a high graft density brush exhibits little change, except near the brush free end (see \fref{numden_PNIPAm}) and results in a very small swelling ratio at $20~^{\circ}C$. The volume fraction in a high density brush is largely close to the minimum in the interaction free energy \emph{vs} volume fraction curve in \fref{Fint_PNIPAm}, even at $20~^{\circ}C$. When temperature rises, thermodynamic stability forbids only smaller volume fractions and not these high values, and also the volume fraction at which the minimum in interaction free energy occurs changes very little. As a result, the profile registers minimal changes on increasing temperature, except near the brush free end where smaller volume fractions occur below $30.5~^{\circ}C$.

The ratio $\frac{\partial\epsilon_{\textsc{zz}}}{\partial\epsilon_{\textsc{xx}}}$ in a brush, which is akin to negative of Poisson's ratio in continuum mechanics, shows qualitatively different behavior even at good solvent condition ($T=20~^{\circ}C$) as graft density is increased from low to moderate to high. In a good solvent at low graft density it is $\approx -1/3$, as predicted by the SST with Gaussian chains\cite{manav2018}. At moderate graft density however, it is negative near the brush free end, but becomes positive close to the grafting surface. For high density brushes, it is positive in a narrow domain close to the brush free end, but approaches $-1$ in the bulk of the brush, signifying little change in volume and hence only a small solvent movement due to the applied strain. In the case of phase separated brushes also it is $\lesssim -1$ in the dense region close to the grafting surface. In contrast to a solid material, where this ratio varies in the range $[-0.5~1]$, in a polymer brush, the minimum value of this ratio is $-1$. On the higher end, it goes far above $1$. This comparison however must be supplemented with the differences in the two systems: in a solid, no atoms or molecules from a representative volume can move in or out, whereas solvent is allowed to move in and out of a layer in a polymer brush.

The stress in a brush has contribution from both the interaction among the monomers and the elastic stretching of the chains. The stress due to the interaction among the monomers is always compressive. The stress due to the elastic stretching is found to be tensile in some sections of a moderate to high density brush at temperatures below $30.5~^{\circ}C$. However, summation of the two is always compressive. For a low graft density swollen brush (at temperatures below transition temperature for the brush), the stress contribution from the interaction among the monomers dominates. As temperature increases, the contribution to the stress from the interaction among the monomers decays. Since brush height in a low graft density brush also decreases with increasing temperature due to accumulation of the monomers close to the grafting surface, the contribution to the stress from the elastic stretching of the chains also decreases but comparatively slowly. As a result, in a collapsed low graft density brush, the stress contribution from the elastic stretching dominates. Also, an overall large decrease in the magnitude of the stress is observed. For high graft density brushes, at all temperatures, the stress contribution from the elastic stretching dominates. Since brush heights do not change appreciably for high density brushes due to a change in temperature, the stress also shows only a small change.

The stress in a brush and the ability to tune them by changing temperature exhibit contrasting behavior. Whereas the resultant stress increases in magnitude with increasing graft density, the stress tuneability ratio decrease. This finding has significant implications for the application of brushes. To get large bending of a flexible substrate, or to bend a stiffer substrate, one needs large resultant stress. If this large resultant stress is gotten by high graft density brushes, the ability to tune the stress and consequently the substrate bending decreases. In many an application, the substrate is desired to go from a bent configuration to a flat configuration as temperature is increased (or vice versa). But, for very high density brushes, a flat configuration may not be achievable. Another way to obtain a large resultant stress is by increasing polymer chain length. Importantly, this does not affect tuneability ratio. However, $\tau_s$ varies linearly with $N$, whereas it shows a nonlinear sharp increase in magnitude with increasing graft density. These trade-offs create a design space from which appropriate graft density and molecular weight of polymer chains can be chosen depending on the application and the fabrication constraints.

Finally, by observing example of the radius of curvature of the deformed beams of different thicknesses, we find that to estimate the deformation of the substrates of thicknesses comparable to brush thickness, knowledge of the stress distribution, which allows calculation of resultant moment along with resultant stress, is imperative.

\section{Conclusion}
\label{conclusion}
In this work, we have developed a semi-analytical strong stretching theory for stimuli-responsive brushes of any arbitrary density using the Langevin force-extension relation for an ideal chain. The theory is used to study the stress distribution in the brushes. It was applied to PNIPAM brushes, which show vertical phase separation within a brush in a narrow range of temperature near the critical temperature for PNIPAm solution starting at a graft density dependent transition temperature. Low to moderate graft density brushes show considerable change in the polymer volume fraction profile as well as in the stress profile within the brush with an increase in temperature from $20~^{\circ}C$ to $33~^{\circ}C$. At $33~^{\circ}C$, the brush is in the collapsed state and undergoes minimal change in the volume fraction and stress as the temperature is increased further. For very high graft density brushes, little change in the profiles is observed in the whole range of temperature studied. Accordingly, brush height decreases with an increase in temperature up to $33~^{\circ}C$, after which it plateaus for small to moderate density brushes. For high density brushes, brush height shows a mild increase again above $33~^{\circ}C$. End density within a brush diverges as a temperature change causes a discontinuity in the polymer volume fraction profile.

Furthermore, the theory predicts that the resultant stress and the resultant moment due to a PNIPAM brush is compressive and increases in magnitude with increasing graft density. For lower graft densities, they decrease in magnitude with increasing temperature up to $33~^{\circ}C$ and plateau after that. However, for very high graft density brushes, the trend is reversed. Stress tuneability ratio, describing the ability to tune resultant stress by varying temperature, is predicted to decrease with increasing graft density in contrast with the magnitude of resultant stress. This has important implications in applications involving brushes.

We have ignored substrate curvature in this study and assumed brushes to be planar. The effect of the curvature in a deformed substrate on the stress in a brush will be studied in the future.

\section*{Acknowledgement}
The authors thank Natural Sciences and Engineering Research Council of Canada (NSERC) for its funding through CREATE (NanoMat program at UBC), Discovery grant, and the Collaborative Health Research project jointly with the Canadian Institute of Health Research (CIHR).


\begin{appendix}
	
\section{SST formulation}
\label{SST_formulation}
Elongation of a freely jointed chain due to an end stretching force $p$ is a function of $p$ and is defined as:
\begin{equation}
e(p)=\frac{\textsc{z}}{Na}, \label{eq2} 
\end{equation}
where $\textsc{z}$ is end to end distance, and $Na$ is contour length of a chain. For a chain in a nonuniform field, elongation needs to be defined locally as stretching force changes along the length of the chain. In this case,
\begin{equation}
e(p)=\frac{1}{a}\frac{d\textsc{z}}{dn}, \label{eq3}
\end{equation}
where $dn$ monomers of a chain are contained in a thin layer of thickness $d\textsc{z}$ at height $\textsc{z}$ as shown in \fref{brush_slit}. $p$ is local elongation force and the static equilibrium of n$-{th}$ segment of the chain yields the following relation between the potential field and stretching force\cite{amoskov94}:
\begin{equation}
\frac{dp}{dn}=\frac{dp}{d\textsc{z}}\frac{d\textsc{z}}{dn}=-\frac{\partial V(\textsc{z})}{\partial \textsc{z}}. \label{eq4}
\end{equation}
By combining \eqref{eq3} and \eqref{eq4}, relation between local elongation force in a chain segment at height $\textsc{z}$ and the potential field is obtained.
\begin{equation}
E(p)=a\int_0^p e(p')dp'=V(\textsc{z}_e)-V(\textsc{z}), \label{eq5}
\end{equation}
where $\textsc{z}_e$ is location of the chain end. $E(p)$ is the complementary stretching energy stored in a portion of a chain of length of a monomer and this function depends on the model of ideal chain. This work uses ideal chain model with force-extension relation given by Langevin function to account for finite extensibility of a polymer chain. Also, the above equation suggest that the stretching force in a chain at height $\textsc{z}$ depends on the location of the chain end $\textsc{z}_e$.

In our monodisperse polymer chain, each chain has $N$ monomers. Hence, $\int_0^N dn=N$. Using \eqref{eq3}, this relation can be expressed as:
\begin{equation}
\int_0^N dn=\int_0^{\textsc{z}_e}\frac{d\textsc{z}}{a e(p)}=N. \label{eq6}
\end{equation}
Notice that using \eqref{eq5}, $p=E^{-1}(V(\textsc{z}_e)-V(\textsc{z}))$, where $E^{-1}$ is inverse function of $E$. Hence, $e(p)=\Lambda(V(\textsc{z}_e)-V(\textsc{z}))$, where $\Lambda(\cdot)=e(E^{-1}(\cdot))$. The constraint in \eqref{eq6} can then be expressed as follows:
\begin{equation}
\int_0^{\textsc{z}_e}\frac{d\textsc{z}}{Na \Lambda(V(\textsc{z}_e)-V(\textsc{z}))}=1. \label{eq7}
\end{equation}
The above integral equation can be solved to obtain $V(\textsc{z})$ once an appropriate model of a polymer chain, that is the form of $e(p)$, for the given polymer is ascertained.

Now, let $g(\textsc{z})$ define the density of chain ends within the brush. Then, the self consistency requirement necessitates the following relation between polymer volume fraction ($\phi(\textsc{z})$) and end density\cite{amoskov94}:
\begin{equation}
\phi(\textsc{z})=v_0\int_{\textsc{z}}^H\frac{g(\textsc{z}_e)}{a\Lambda(V(\textsc{z}_e)-V(\textsc{z}))}d\textsc{z}_e, \label{eq9}
\end{equation}
Once $V(\textsc{z})$ is determined by solving the integral equation \eqref{eq7}, $\phi(\textsc{z})$ can be obtained by solving \eqref{eq1}. The above integral equation can then be solved to obtain end density $g(\textsc{z})$ in a brush.

We nondimensionalize the system of equations using the following scheme:
\begin{eqnarray}
\bar{\textsc{z}}=\frac{\textsc{z}}{Na}, \quad \bar{H}=\frac{H}{Na}, \quad \bar{V}(\bar{\textsc{z}})=\frac{V(\textsc{z})}{k_BT}, \quad \bar{\Lambda}(\bar{V}(\bar{\textsc{z}}_e)-\bar{V}(\bar{\textsc{z}}))=\Lambda(V(\textsc{z}_e)-V(\textsc{z})),\nonumber \\ \quad \bar{\phi}(\bar{\textsc{z}})=\phi(\textsc{z}), \quad \bar{g}(\bar{\textsc{z}}')=Nv_0g(\textsc{z}'), \quad \bar{\mu}(\bar{\phi}(\bar{\textsc{z}}))=\frac{\mu(\phi(\textsc{z}))}{k_BT},\quad \bar{p}=\frac{pr_ka}{k_BT},~{\rm and} \quad \bar{e}(\bar{p})=e(p), 
\label{eq7a}
\end{eqnarray}
where $k_B$ is the Boltzmann constant and $T$ is temperature in Kelvin. Then, Eqs. \eqref{eq5}, \eqref{eq7}, and \eqref{eq9} turn into \eqref{eq5b}, \eqref{eq7b}, and \eqref{eq9a}, respectively.

\section{End density distribution}
\label{calc_endden}
At first, polymer volume fraction $\phi(\textsc{z})$ for a brush is calculated using \eqref{eq1}. Then for known $\phi(\textsc{z})$, the integral equation \eqref{eq9a} needs to be solved to find end density distribution. Following the approach employed by Amoskov et al.\cite{amoskov94}, we use the method of Laplace transform to solve the equation.

Starting with \eqref{eq7b}, let us assume $u=\bar{V}(\bar{\textsc{z}}_e)$ and $v=\bar{V}(\bar{\textsc{z}})$. Using a change in variable, \eqref{eq7b} can be written as:
\begin{equation}
\int_0^u\frac{d\bar{\textsc{z}}}{dv}\frac{dv}{\bar{\Lambda}(u-v)}=1. \label{eqa1}
\end{equation}
On using Laplace transform for convolution of two functions, the above  transforms to:
\begin{equation}
\mathscr{L}\left(\frac{d\bar{\textsc{z}}_e}{du}\right)\mathscr{L}\left(\frac{1}{\bar{\Lambda}(u)}\right)=\frac{1}{s}. \label{eqa2}
\end{equation}
Then the potential field $u$ can be found by wring the above as
\begin{equation}
\mathscr{L}\left(\frac{d\bar{\textsc{z}}_e}{du}\right)=\frac{1}{s}\left(\mathscr{L}\left(\frac{1}{\bar{\Lambda}(u)}\right)\right)^{-1}, \label{eqa3}
\end{equation}
and obtaining inverse Laplace transform. For a Langevin chain, a series solution for $\bar{V}(\bar{\textsc{z}})$ was obtained\cite{amoskov94} using the above and a Pad\'{e} approximation of the series solution was also reported\cite{biesheuvel2008}.

Depending on $\bar{\phi}(\bar{\textsc{z}})$ profile, we may have distinct cases to solve to obtain end density profile. Here, we consider two cases. In the first case, there are no discontinuities in the polymer volume fraction profile inside the brush, though it can have a discontinuity at the brush free end. In the second case, the brush has a discontinuity in the polymer volume fraction profile inside the brush. Below, we first obtain the expression for end distribution $\bar{g}(\bar{\textsc{z}})$ for the first case, followed by the second case.

\subsection{No jump in the volume fraction profile inside the brush}
\label{endden_nojump}
Let us assume $\bar{u}=\bar{V}(\bar{H})-\bar{V}(\bar{\textsc{z}})$ and $\bar{v}=\bar{V}(\bar{H})-\bar{V}(\bar{\textsc{z}}_e)$. Then \eqref{eq9a} transforms to the following:
\begin{equation}
R(\bar{u})=\int_0^{\bar{u}}\frac{q(\bar{v})}{\bar{\Lambda}(\bar{u}-\bar{v})}d\bar{v}, \quad q(\bar{v})=-\bar{g}(\bar{\textsc{z}}_e)\frac{d\bar{\textsc{z}}_e}{d\bar{v}}, \quad R(\bar{u})=\bar{\phi}(\bar{\textsc{z}}). \label{eqa4}
\end{equation}
By taking the Laplace transform of the above, we obtain:
\begin{equation}
\mathscr{L}(R(\bar{u}))=\mathscr{L}(q(\bar{u}))\mathscr{L}\left(\frac{1}{\bar{\Lambda}(\bar{u})}\right). \label{eqa5}
\end{equation}
Using \eqref{eqa2} in the above, we obtain:
\begin{align}
\mathscr{L}(q(\bar{u}))=&s\mathscr{L}(R(\bar{u}))\mathscr{L}\left(\frac{d\bar{\textsc{z}}_f}{d\bar{u}}\right) \nonumber \\
=&\left(\mathscr{L}\left(\frac{dR(\bar{u})}{d\bar{u}}\right)+R(0) \right)\mathscr{L}\left(\frac{d\bar{\textsc{z}}_f}{d\bar{u}}\right), \label{eqa6}
\end{align}
where $\bar{u}=\bar{V}(\bar{H})-\bar{V}(\bar{\textsc{z}})=\bar{V}(\bar{\textsc{z}}_f)$. By taking inverse Laplace transform of the above equation, we obtain:
\begin{equation}
q(\bar{u})=\int_0^{\bar{u}}\frac{d\bar{\textsc{z}}'_f}{d\bar{v}} \frac{dR(\hat{v})}{d\hat{v}}d\bar{v}+R(0)\frac{d\bar{\textsc{z}}_f}{d\bar{u}}, \label{eqa7}
\end{equation}
where $\hat{v}=\bar{u}-\bar{v}$ and $\bar{v}=\bar{V}(\bar{H})-\bar{V}(\bar{\textsc{z}}_e)=\bar{V}(\bar{\textsc{z}}'_f)$. Then using \eqref{eqa4}, we obtain the expression for end density.
\begin{equation}
\bar{g}(\bar{\textsc{z}})=\frac{d\bar{V}(\bar{\textsc{z}})}{d\bar{\textsc{z}}}\left(\int_0^{\bar{\textsc{z}}_f}\frac{dR(\hat{v})}{d\hat{v}}d\bar{\textsc{z}}'_f+R(0)\frac{d\bar{\textsc{z}}_f}{d\bar{u}} \right). \label{eqa8}
\end{equation}
As $R(0)=\bar{\phi}(\bar{H})$, the last term contributes only when the volume fraction profile has a jump at the brush free end. Also, as $\bar{\textsc{z}} \to \bar{H}$, $\bar{\textsc{z}}_f \to 0$ and $\frac{d\bar{u}}{d\bar{\textsc{z}}_f}\to 0$. Thus, end density diverges and has a vertical asymptote at $\bar{\textsc{z}}=\bar{H}$ when $\bar{\phi}(\bar{H})\ne 0$, as observed in a collapsed brush.

To calculate $\frac{dR(\hat{v})}{d\hat{v}}$, we write $\hat{v}=\bar{u}-\bar{v}=\bar{V}(\bar{H})-\bar{V}(\bar{\textsc{z}})-\bar{V}(\bar{\textsc{z}}'_f)=\bar{V}(\bar{H})-\bar{V}(\bar{\textsc{z}}^*_f)$, where $\bar{V}(\bar{\textsc{z}})+\bar{V}(\bar{\textsc{z}}'_f)=\bar{V}(\bar{\textsc{z}}^*_f)$. Also, $R(\hat{v})=\bar{\phi}(\bar{\textsc{z}}^*_f)$ using \eqref{eqa4}. Then,
\begin{equation}
\frac{dR(\hat{v})}{d\hat{v}}=\frac{d\bar{\phi}(\bar{\textsc{z}}^*_f)}{d\hat{v}}=-\frac{d\bar{\phi}(\bar{\textsc{z}}^*_f)}{d\bar{V}(\bar{\textsc{z}}^*_f)}, \label{eqa8a}
\end{equation}
where $\bar{\phi}(\bar{\textsc{z}}^*_f)$ is obtained by using \eqref{eq1} as follows:
\begin{equation}
\bar{V}(\bar{H})-\bar{V}(\bar{\textsc{z}}^*_f)=\bar{V}(\bar{H})-\bar{V}(\bar{\textsc{z}})-\bar{V}(\bar{\textsc{z}}'_f)=\mu(\bar{\phi}(\bar{\textsc{z}}^*_f))-\bar{\mu}(\bar{\phi}(\bar{H}))). \label{eqa8b}
\end{equation}
Furthermore, differentiation of the above equation yields:
\begin{equation}
-\frac{d\bar{V}(\bar{\textsc{z}}^*_f)}{d\bar{\phi}(\bar{\textsc{z}}^*_f)}=\frac{d\bar{\mu}(\bar{\phi}(\bar{\textsc{z}}^*_f))}{d\bar{\phi}(\bar{\textsc{z}}^*_f)}. \label{eqa8c}
\end{equation}
Since $\frac{d\bar{\phi}(\bar{\textsc{z}}^*_f)}{d\bar{V}(\bar{\textsc{z}}^*_f)}=\left(\frac{d\bar{V}(\bar{\textsc{z}}^*_f)}{d\bar{\phi}(\bar{\textsc{z}}^*_f)} \right)^{-1}$, and $\bar{\mu}(\bar{\phi})$ and consequently its derivative are known for the polymer, the above equation allows evaluation of \eqref{eqa8a}.

\subsection{Jump in the volume fraction profile inside the brush}
\label{endden_jump}
If the polymer volume fraction profile has a jump at the height $\bar{\textsc{z}}=\bar{H}_t$ inside the brush, then \eqref{eqa4} needs to be divided into the two following cases:
\begin{equation}
R(\bar{u})=\begin{cases}
\int_0^{\bar{u}}\frac{q(\bar{v})}{\bar{\Lambda}(\bar{u}-\bar{v})}d\bar{v}, & \bar{u}<\bar{u}_t\\
\int_0^{\bar{u}_t-}\frac{q(\bar{v})}{\bar{\Lambda}(\bar{u}-\bar{v})}d\bar{v}+v_0\int_{\bar{u}_t+}^{\bar{u}}\frac{q(\bar{v})}{\bar{\Lambda}(\bar{u}-\bar{v})}d\bar{v}, & \bar{u}>\bar{u}_t
\end{cases}
\label{eqa9}
\end{equation}
where $\bar{u}|_{\textsc{z}=H_t}=\bar{u}_t$, and $\bar{u}_t-=\bar{u}_t-\epsilon$ and $\bar{u}_t+=\bar{u}_t+\epsilon$, $\epsilon \to 0$. In the first case, that is when $\bar{H}_t<\bar{\textsc{z}}<\bar{H}$, end density is given by \eqref{eqa8}. In this section, we obtain the end density for the second case, that is $\bar{\textsc{z}}<\bar{H}_t$.

In the second case, the first integral in the expression can be calculated by using the result from the first case. Then the equation can be transformed into the following form:
\begin{equation}
S(\tilde{u})=\int_0^{\tilde{u}}\frac{\tilde{q}(\tilde{v})}{\bar{\Lambda}(\tilde{u}-\tilde{v})}d\tilde{v}, \label{eqa10}
\end{equation}
where, 
\begin{align}
S(\tilde{u})&=R(\bar{u})-\int_0^{\bar{u}_t-}\frac{q(\bar{v}_h)}{\bar{\Lambda}(\bar{u}-\bar{v}_h)}d\bar{v}_h \nonumber \\
&=R(\tilde{u}+\alpha)-\int_0^{\bar{u}_t-}\frac{q(\bar{v}_h)}{\bar{\Lambda}(\tilde{u}+\alpha-\bar{v}_h)}d\bar{v}_h, \label{eqa11}
\end{align}
and $\tilde{u}=\bar{V}(\bar{H}_t)-\bar{V}(\bar{\textsc{z}})=\bar{u}-\alpha$, $\alpha=\bar{V}(\bar{H})-\bar{V}(\bar{H}_t)$, $\tilde{v}=\bar{V}(\bar{H}_t)-\bar{V}(\bar{\textsc{z}}_e)$, $\bar{v}_h=\bar{V}(\bar{H})-\bar{V}(\bar{\textsc{z}}_h)$,  $q(\bar{v})d\bar{v}=\tilde{q}(\tilde{v})d\tilde{v}=-\bar{g}(\bar{\textsc{z}}_e)d\bar{\textsc{z}}_e$, and we have replaced the integration variable $\bar{v}$ with $\bar{v}_h$. Note that \eqref{eqa10} has the same form as \eqref{eqa4}. So, following the same procedure as employed in solving \eqref{eqa4}, we obtain:
\begin{equation}
\tilde{q}(\tilde{u})=\int_0^{\tilde{u}}\frac{d\bar{\textsc{z}}'_g}{d\tilde{v}}\frac{dS(\check{v})}{d\check{v}}d\tilde{v}+S(0)\frac{d\bar{\textsc{z}}_g}{d\tilde{u}}, \label{eqa12}
\end{equation}
where $\check{v}=\tilde{u}-\tilde{v}$, and $\tilde{u}=\bar{V}(\bar{H}_t)-\bar{V}(\bar{\textsc{z}})=\bar{V}(\bar{\textsc{z}}_g)$ and $\tilde{v}=\bar{V}(\bar{H}_t)-\bar{V}(\bar{\textsc{z}}_e)=\bar{V}(\bar{\textsc{z}}'_g)$. Then using the relation $\bar{g}(\bar{\textsc{z}})=-\tilde{q}(\tilde{u})\frac{d\tilde{u}}{d\bar{\textsc{z}}}$, we obtain the expression for $\bar{g}(\bar{\textsc{z}})$.
\begin{equation}
\bar{g}(\bar{\textsc{z}})=\frac{d\bar{V}(\bar{\textsc{z}})}{d\bar{\textsc{z}}}\left(\int_0^{\bar{\textsc{z}}_g}\frac{dS(\check{v})}{d\check{v}}d\bar{\textsc{z}}'_g+S(0)\frac{d\bar{\textsc{z}}_g}{d\tilde{u}}\right). \label{eqa13}
\end{equation}
Note that $\tilde{u}=0$, when $\bar{\textsc{z}}=\bar{H}_t$. Hence, $S(0)=\bar{\phi}(\bar{H}_t-)-\bar{\phi}(\bar{H}_t+)$, that is, it equals the jump in the polymer volume fraction at $\bar{\textsc{z}}=\bar{H}_t$. Also, since $\bar{\textsc{z}}_g \to 0$ and $\frac{d\tilde{u}}{d\bar{\textsc{z}}_g}\to 0$ when $\bar{\textsc{z}} \to \bar{H}_t$, end density diverges and has a vertical asymptote at $\bar{\textsc{z}}=\bar{H}_t$ within the brush.

To be able to calculate $\bar{g}(\bar{\textsc{z}})$, we first need to determine $\frac{dS(\check{v})}{d\check{v}}$. By replacing $\tilde{u}$ with $\check{v}$ in \eqref{eqa11}, and taking derivative, we get:
\begin{align}
\frac{dS(\check{v})}{d\check{v}}&=\frac{dR(\check{v}+\alpha)}{d\check{v}}+\int_0^{\bar{u}_t-}\frac{q(\bar{v}_h)}{(\bar{\Lambda}(\check{v}+\alpha-\bar{v}_h))^2}\frac{d\bar{\Lambda}(\check{v}+\alpha-\bar{v}_h)}{d\check{v}}d\bar{v}_h \nonumber \\
&=\frac{dR(\check{v}+\alpha)}{d\check{v}}+\int_{\bar{H}_t}^{\bar{H}}\frac{\bar{g}(\bar{\textsc{z}}_h)d\bar{\textsc{z}}_h}{(\bar{\Lambda}(\check{v}+\alpha-\bar{v}_h))^2}\frac{d\bar{\Lambda}(\check{v}+\alpha-\bar{v}_h)}{d\check{v}}. \label{eqa14}
\end{align}
To calculate $\frac{dR(\check{v}+\alpha)}{d\check{v}}$, we write $\check{v}+\alpha=\bar{V}(\bar{H})-\bar{V}(\bar{\textsc{z}})-\bar{V}(\bar{\textsc{z}}'_g)=\bar{V}(\bar{H})-\bar{V}(\bar{\textsc{z}}^*_g)$, where $\bar{V}(\bar{\textsc{z}})+\bar{V}(\bar{\textsc{z}}'_g)=\bar{V}(\bar{\textsc{z}}^*_g)$. Then $R(\check{v}+\alpha)=\bar{\phi}(\bar{\textsc{z}}^*_g)$. Also,
\begin{equation}
\frac{dR(\check{v}+\alpha)}{d\check{v}}=\frac{d\bar{\phi}(\bar{\textsc{z}}^*_g)}{d\check{v}}=-\frac{d\bar{\phi}(\bar{\textsc{z}}^*_g)}{d\bar{V}(\bar{\textsc{z}}^*_g)}. \label{eqa15}
\end{equation}
$\bar{\phi}(\bar{\textsc{z}}^*_g)$ is obtained by solving the following equation obtained by using \eqref{eq1}.
\begin{equation}
\bar{V}(\bar{H})-\bar{V}(\bar{\textsc{z}}^*_g)=\bar{V}(\bar{H})-\bar{V}(\bar{\textsc{z}})-\bar{V}(\bar{\textsc{z}}'_g)=\bar{\mu}(\bar{\phi}(\bar{\textsc{z}}^*_g))-\bar{\mu}(\bar{\phi}(\bar{H}))). \label{eqa16}
\end{equation}
Afterwards differentiation of the above relation gives:
\begin{equation}
-\frac{d\bar{V}(\bar{\textsc{z}}^*_g)}{d\bar{\phi}(\bar{\textsc{z}}^*_g)}=\frac{d\bar{\mu}(\bar{\phi}(\bar{\textsc{z}}^*_g))}{d\bar{\phi}(\bar{\textsc{z}}^*_g)}. \label{eqa17}
\end{equation}
As $\frac{d\bar{\phi}(\bar{\textsc{z}}^*_g)}{d\bar{V}(\bar{\textsc{z}}^*_g)}=\left(\frac{d\bar{V}(\bar{\textsc{z}}^*_g)}{d\bar{\phi}(\bar{\textsc{z}}^*_g)} \right)^{-1}$, $\frac{dR(\check{v}+\alpha)}{d\check{v}}$ can be calculated.

Evaluation of the term with integral in \eqref{eqa14} requires us to first calculate $\frac{d\bar{\Lambda}(\check{v}+\alpha-\bar{v}_h)}{d\check{v}}$ which can be simplified to the following form:
\begin{equation}
\frac{d\bar{\Lambda}(\check{v}+\alpha-\bar{v}_h)}{d\check{v}}=-\frac{d\bar{\Lambda}(\bar{V}(\bar{\textsc{z}}_h)-\bar{V}(\bar{\textsc{z}}^*_g))}{d\bar{V}(\bar{\textsc{z}}^*_g)}. \label{eqa18}
\end{equation}
Remember that $\bar{\Lambda}(\bar{V}(\bar{\textsc{z}}_h)-\bar{V}(\bar{\textsc{z}}^*_g))=\bar{e}(\bar{p}^*_g)$, where $\bar{p}^*_g$  is the local stretching force at height $\bar{\textsc{z}}^*_g$ in a polymer chain with end at $\bar{\textsc{z}}_h$, and $\bar{e}(\bar{p}^*_g)$ is local stretching in a chain due to the stretching force $\bar{p}^*_g$. Also, from \eqref{eq5b}, $\bar{E}(\bar{p}^*_g)=\bar{V}(\bar{\textsc{z}}_h)-\bar{V}(\bar{\textsc{z}}^*_g)$. Taking derivative of this expression gives:
\begin{equation}
\frac{1}{r_k}\bar{e}(\bar{p}^*_g)\frac{d \bar{p}^*_g}{d\bar{V}(\bar{\textsc{z}}^*_g)}=-1, \label{eqa19}
\end{equation}
where we have used $\frac{d\bar{E}(\bar{p}^*_g)}{d\bar{p}^*_g}=\frac{\bar{e}(\bar{p}^*_g)}{r_k}$ following \eqref{eq5b}. Then substituting the above result in \eqref{eqa18} yields:
\begin{equation}
\frac{d\bar{\Lambda}(\check{v}+\alpha-\bar{v}_h)}{d\check{v}}=-\frac{d\bar{\Lambda}(\bar{V}(\bar{\textsc{z}}_h)-\bar{V}(\bar{\textsc{z}}^*_g))}{d\bar{V}(\bar{\textsc{z}}^*_g)}=-\frac{d\bar{e}(\bar{p}^*_g)}{d\bar{p}^*_g}\frac{d\bar{p}^*_g}{d\bar{V}(\bar{\textsc{z}}^*_g)}=\frac{r_k}{\bar{e}(\bar{p}^*_g)}\frac{d\bar{e}(\bar{p}^*_g)}{d\bar{p}^*_g}. \label{eqa20}
\end{equation}

\section{Stress calculation}
\label{stress_calc}
Stress calculation in the regions with continuous volume fraction profile is described first in this section, followed by calculation of stress at the interface.

\subsection{In the regions with continuous volume fraction}
Evaluating stress requires us to first calculate the derivative terms $\frac{\partial\epsilon_{\textsc{zz}}(\bar{\textsc{z}})}{\partial\epsilon_{\textsc{xx}}}$ and $\frac{\partial f(\bar{\textsc{z}})}{\partial\epsilon_{\textsc{xx}}}$ in \eqref{eq18}. We start by calculating $\frac{\partial\epsilon_{\textsc{zz}}(\bar{\textsc{z}})}{\partial\epsilon_{\textsc{xx}}}$, which can be written as:
\begin{equation}
\frac{\partial\epsilon_{\textsc{zz}}}{\partial\epsilon_{\textsc{xx}}}=\frac{\partial}{\partial \epsilon_{\textsc{xx}}}\left(\frac{\partial \bar{u}_{\textsc{z}}}{\partial \bar{\textsc{z}}}\right)=\frac{\partial}{\partial \bar{\textsc{z}}}\left(\frac{\partial \bar{u}_{\textsc{z}}}{\partial \epsilon_{\textsc{xx}}}\right),
\label{eq19}
\end{equation} 
where $\bar{u}_{\textsc{z}}=u_{\textsc{z}}/(Na)$ is the nondimensionalized displacement of a thin layer at $\textsc{z}$ shown in~\fref{brush_slit} in the $\textsc{z}$-direction due to the applied strain $\epsilon_{\textsc{xx}}$. In order to evaluate $\frac{\partial \bar{u}_{\textsc{z}}}{\partial \epsilon_{\textsc{xx}}}$ and subsequently $\frac{\partial\epsilon_{\textsc{zz}}}{\partial\epsilon_{\textsc{xx}}}$, we enforce the condition that a a thin layer within the brush as shown in~\fref{brush_slit} is defined by the monomers (total number $V_0\bar{\phi}/v_0$, where $V_0$ is initial layer volume) inside it, and on applying a strain, no monomer moves in or out of the layer. Hence, $\Delta(V_0\bar{\phi}/v_0)=0$, which on simplification yields~\citep{manav2018}:
\begin{equation}
\frac{\partial}{\partial \bar{\textsc{z}}}\left(\frac{\partial \bar{u}_{\textsc{z}}}{\partial \epsilon_{\textsc{xx}}}\right)=-\frac{1}{\bar{\phi}}\frac{\partial \bar{\phi}}{\partial \epsilon_{\textsc{xx}}}-1,
\label{eq20}
\end{equation}
with the boundary condition in the case of no vertical phase separation within a brush as given below:
\begin{equation}
\left[\frac{\partial \bar{u}_{\textsc{z}}}{\partial \epsilon_{\textsc{xx}}} \right]_{\bar{\textsc{z}}=\bar{H}}=\frac{\partial \bar{H}}{\partial \epsilon_{\textsc{xx}}}.
\label{eq21}
\end{equation}
We obtain $\frac{\partial \bar{\phi}}{\partial \epsilon_{\textsc{xx}}}$ by taking derivative of the \eqref{eq1} and simplifying the resulting expression.
\begin{equation}
\frac{\partial \bar{\phi}}{\partial \epsilon_{\textsc{xx}}}=\left(\frac{d\bar{V}(\bar{H})}{d\bar{H}}\frac{\partial \bar{H}}{\partial \epsilon_{\textsc{xx}}}-\frac{d\bar{V}(\bar{\textsc{z}})}{d\bar{\textsc{z}}}\frac{\partial \bar{u}_{\textsc{z}}}{\partial \epsilon_{\textsc{xx}}} \right)\frac{d\bar{\mu}(\bar{\phi})}{d\bar{\phi}},
\label{eq22}
\end{equation}
where $\frac{\partial \bar{H}}{\partial \epsilon_{\textsc{xx}}}=\frac{\partial \bar{H}}{\partial \rho_g}\frac{d\rho_g}{d\epsilon_{\textsc{xx}}}$ is obtained by  numerically solving \eqref{eq1} for a range of graft densities and obtaining $\frac{\partial \bar{H}}{\partial \rho_g}$. Since graft density after the strain $\epsilon_{\textsc{xx}}$ is applied is  $\approx\rho_{g}(1-\epsilon_{\textsc{xx}})$, $\frac{d\rho_{g}}{d\epsilon_{\textsc{xx}}}=-\rho_g$. After that, by solving \eqref{eq20} along with \eqref{eq22} and the boundary condition in \eqref{eq21}, $\frac{\partial \bar{u}_{\textsc{z}}}{\partial \epsilon_{\textsc{xx}}}$ is obtained.

In the case of a vertical phase separation within a brush, domain is divided into two. For the inner domain, that is $\bar{\textsc{z}}<\bar{H}_t$, boundary condition is as follows:
\begin{equation}
\left[\frac{\partial \bar{u}_{\textsc{z}}}{\partial \epsilon_{\textsc{xx}}} \right]_{\bar{\textsc{z}}=0}=0.
\label{eq23}
\end{equation}
For $\bar{H}_t<\bar{\textsc{z}}\le\bar{H}$, \eqref{eq20} is solved with the following boundary condition:
\begin{equation}
\left[\frac{\partial \bar{u}_{\textsc{z}}}{\partial \epsilon_{\textsc{xx}}} \right]_{\bar{\textsc{z}}=\bar{H}}=\frac{\partial \bar{H}}{\partial \epsilon_{\textsc{xx}}}.
\label{eq24}
\end{equation}

It should be noted that this leads to a discontinuous $\bar{u}_{\textsc{z}}(\bar{\textsc{z}})$ and consequently discontinuous $\frac{\partial\epsilon_{\textsc{zz}}}{\partial\epsilon_{\textsc{xx}}}$. In a solid material, a discontinuous $\bar{u}_{\textsc{z}}$ results from fracture. However, here a discontinuous $\bar{u}_{\textsc{z}}$ indicates that in a layer very close to $\bar{H}_t$, polymer volume fraction undergoes a sudden jump as a result of the applied strain $\epsilon_{\textsc{xx}}$.

To calculate the derivative of free energy density, the derivative of the two contributions to the free energy density are calculated separately.
\begin{equation}
\frac{\partial f(\bar{\textsc{z}})}{\partial\epsilon_{\textsc{xx}}}=\frac{\partial f_{int}(\bar{\textsc{z}})}{\partial\epsilon_{\textsc{xx}}}+\frac{\partial f_{el}(\bar{\textsc{z}})}{\partial\epsilon_{\textsc{xx}}}. \label{eq25}
\end{equation}
To calculate the first term, we recognize that
\begin{equation}
\frac{\partial f_{int}(\bar{\textsc{z}})}{\partial\epsilon_{\textsc{xx}}}=\frac{k_BT}{v_0}\bar{\mu}(\bar{\phi}(\bar{\textsc{z}}))\frac{\partial \bar{\phi}(\bar{\textsc{z}})}{\partial\epsilon_{\textsc{xx}}}, \label{eq26}
\end{equation}
and $\frac{\partial \bar{\phi}(\bar{\textsc{z}})}{\partial\epsilon_{\textsc{xx}}}$ is obtained from \eqref{eq22}. To evaluate $\frac{\partial f_{el}(\bar{\textsc{z}})}{\partial\epsilon_{\textsc{xx}}}$, we take derivative of \eqref{eq15}, which yields:
\begin{align}
\frac{\partial f_{el}}{\partial \epsilon_{\textsc{xx}}}=&\frac{k_BT}{v_0r_k}\int_{\bar{\textsc{z}}}^{\bar{H}}\log\left(\frac{\sinh(\bar{p})}{\bar{p}} \right)\left(\frac{1-(\coth(\bar{p}))^2+\frac{1}{\bar{p}^2}}{(\bar{e}(\bar{p}))^2}\right)\bar{g}(\bar{\textsc{z}}_e)\frac{\partial \bar{p}}{\partial \epsilon_{\textsc{xx}}}d\bar{\textsc{z}}_e \nonumber \\
&+\frac{k_BT}{v_0r_k}\int_{\bar{\textsc{z}}}^{\bar{H}} \left(\bar{p}-\frac{1}{\bar{e}(\bar{p})}\log\left(\frac{\sinh(\bar{p})}{\bar{p}} \right)\right) \frac{\partial \bar{g}(\bar{\textsc{z}}_e)}{\partial \epsilon_{\textsc{xx}}}d\bar{\textsc{z}}_e, \label{eq27}
\end{align}
where
\begin{equation}
\frac{\partial \bar{p}}{\partial \epsilon_{\textsc{xx}}}=-\frac{r_k}{\bar{e}(\bar{p})}\frac{\partial \bar{V}(\bar{\textsc{z}})}{\partial \epsilon_{\textsc{xx}}}=-\frac{1}{\bar{e}(\bar{p})}\frac{8}{5}\bar{\textsc{z}}\frac{2\bar{\textsc{z}}^4-4\bar{\textsc{z}}^2+5}{(1-\bar{\textsc{z}}^2)^2}\frac{\partial \bar{u}_{\textsc{z}}}{\partial \epsilon_{\textsc{xx}}}. \label{eq28}
\end{equation}
Since $\bar{e}(\bar{p})\to 0$ as $\bar{p}\to 0$, $\frac{\partial \bar{p}}{\partial \epsilon_{\textsc{xx}}}\to \infty$. Also, in the case of phase separation, $\bar{g}(\bar{\textsc{z}}_e)$ diverges. These lead to improper integrals in the expression for $\frac{\partial f_{el}}{\partial \epsilon_{\textsc{xx}}}$ which require appropriate variable transformations in the expression for $f_{el}$ in order to numerically calculate them.

\subsection{At the interface}
The free energy densities can be divided in their interaction and elastic components, which can be written as follows using \eqref{eq14} and \eqref{eq15}.
\begin{align}
f_{1int}&=\left[(1-\phi^+)\log(1-\phi^+)+\chi \phi^+(1-\phi^+)\right]\frac{k_BT}{v_0}, \nonumber \\
f_{2int}&=\left[(1-\phi^-)\log(1-\phi^-)+\chi \phi^-(1-\phi^-)\right]\frac{k_BT}{v_0}, \nonumber \\
f_{1el}&=\frac{k_BT}{v_0}\int_{\bar{H}_t}^{\bar{H}}\frac{1}{r_k\bar{e}(\bar{p})}\left(\bar{p}\bar{e}(\bar{p})-\log\left(\frac{\sinh(\bar{p})}{\bar{p}} \right)\right)\bar{g}(\bar{\textsc{z}}_e)d\bar{\textsc{z}}_e, \nonumber \\
f_{2el}&=f_{1el}+\left[\frac{\partial f_{el}(\bar{\textsc{z}})}{\partial\epsilon_{\textsc{xx}}}\right]_{\bar{\textsc{z}}=\bar{H}_t}\delta\epsilon_{\textsc{xx}},
\label{eq19g}
\end{align}
where the expression for $f_{2el}$ uses continuity of $f_{el}$ even in case of vertical phase separation. Note that the discontinuity in volume fraction causes only a corner in elastic free energy density profile. $f_{1el}$ and $\left[\frac{\partial f_{el}(\bar{\textsc{z}})}{\partial\epsilon_{\textsc{xx}}}\right]_{\bar{\textsc{z}}=\bar{H}_t}$ have already been calculated in the previous section. Substituting \eqref{eq19g} in \eqref{eq19f} yields:
\begin{equation}
\sigma_{\textsc{xx}}(\bar{H}_t)=\lim_{\delta\epsilon_{\textsc{xx}}\to 0}\frac{1}{\phi^-}\frac{(f_{2int}\phi^+-f_{1int}\phi^-)+f_{1el}(\phi^+-\phi^-)}{\delta\epsilon_{\textsc{xx}}}+\frac{\phi^+}{\phi^-}\left[\frac{\partial f_{el}(\bar{\textsc{z}})}{\partial\epsilon_{\textsc{xx}}}\right]_{\bar{\textsc{z}}=\bar{H}_t}.
\label{eq19h}
\end{equation}

\section{Brush height and swelling ratio}
\label{H_SR}
\begin{figure}[!h]
	\center
	\includegraphics[width=8cm]{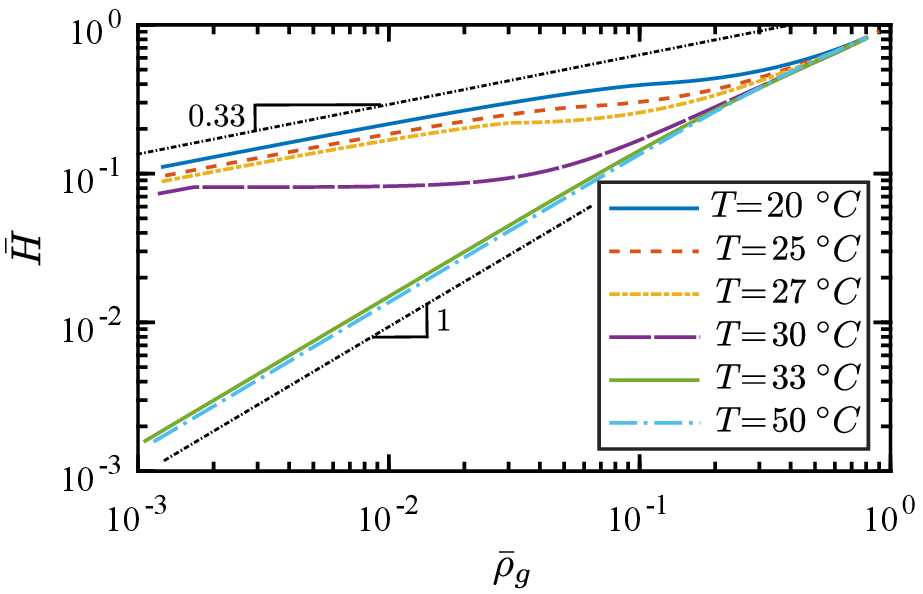}
	\hspace{-8mm}
	\includegraphics[width=8cm]{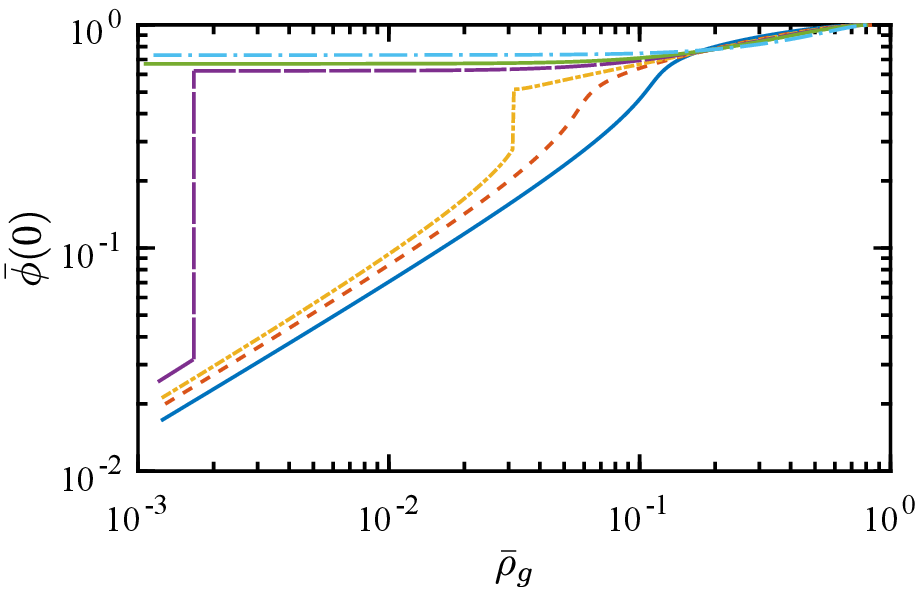}
	\caption{Variation of brush height and polymer volume fraction at $\bar{\textsc{z}}=0$ with graft density at different temperatures. Notice that temperature has little effect on the brush height at very high graft densities.}
	\label{H_rhog_Tvary_PNIPAm}
\end{figure}

Variation in brush height with graft density is also temperature dependent ($\bar{H} \sim \rho_g^{n(T,\rho_g)}$) and shows a nonmonotonic behavior in a narrow temperature range as shown in \fref{H_rhog_Tvary_PNIPAm}. At temperatures below $30~^{\circ}C$, $n\approx 1/3$ in the low graft density regime. At very high graft densities, $n$ approaches $1$, as expected for extremely dense brushes. In between these two limits, $n$ exhibits atypical behavior and approaches $0$ before recovering and approaching $1$. At $33~^{\circ}C$ and above, the height variation is close to $\bar{\rho}_g$. 

Maximum polymer volume fraction in a brush occurs near the grafting surface and is plotted as a function of graft density in \fref{H_rhog_Tvary_PNIPAm} at different temperatures. Like the volume fraction profile, it shows a discontinuity at temperatures between $26.36~^{\circ}C$ and $30.5~^{\circ}C$. At temperatures above $30.5~^{\circ}C$, it shows minimal change with graft density for low graft density brushes.

\begin{figure}[!h]
	\center
	\includegraphics[width=8.5cm]{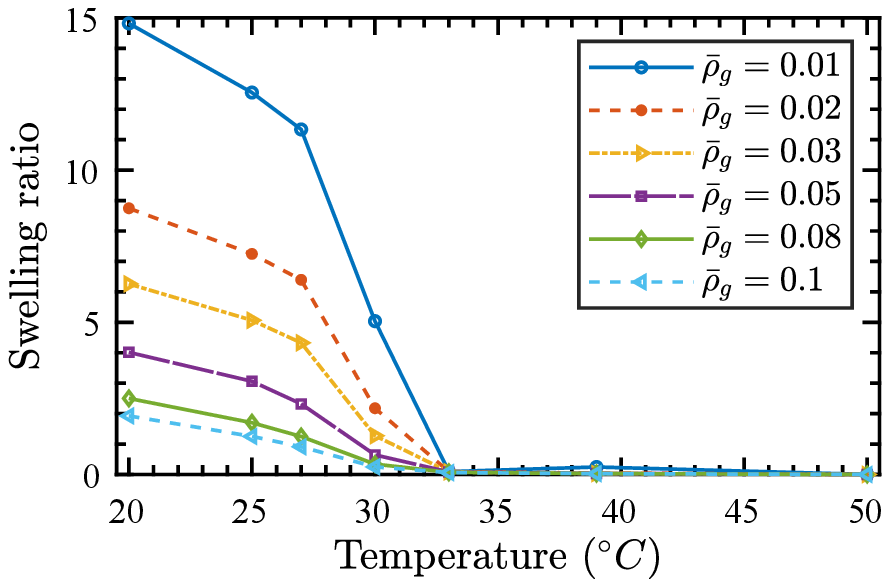}
	\hspace{-8mm}
	\includegraphics[width=8.5cm]{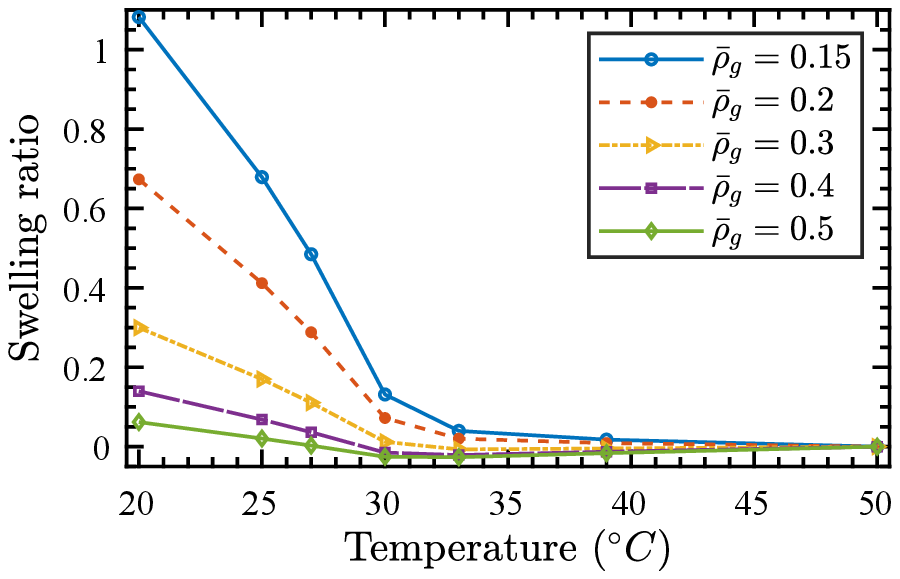}
	\caption{Swelling ratio \emph{vs} temperature plot for different graft densities. It decreases with increasing temperature monotonically, except for very high graft densities ($\bar{\rho}_g>0.2$).}
	\label{swellingRatio_T_rhogvary_PNIPAm}
\end{figure}

Swelling ratio, defined here as $\bar{H}_{T}/\bar{H}_{50^{\circ}C}-1$, where $\bar{H}_T$ is the brush height at temperature $T$, is plotted as a function of temperature for different graft densities in \fref{swellingRatio_T_rhogvary_PNIPAm}. 
Swelling ratio decreases with increasing graft density and is very small at very high graft densities. Also, it decreases with an increase in temperature up to $33~^{\circ}C$ and plateaus above this temperature for $\bar{\rho}_g\le0.2$. However, it shows a nonmonotonic behavior for $\bar{\rho}_g>0.2$. Malham et al.\cite{malham2010}, and Witte et al.\cite{witte2020} have reported that swelling ratio of a PNIPAm brush smoothly decrease with increasing temperature, and increases with decreasing graft density. Devaux et al.\cite{devaux2005} have reported maximum swelling ratio of $0.2-0.4$ for polystyrene brushes with maximum volume fraction up to 0.85. Our result is consistent with these experimental measurements.

Interestingly, dependence of brush height on graft density even in a good solvent condition diminishes ($n(T,\bar{\rho}_g)<1/3$) as graft density increases, before picking up again above $\bar{\rho}_g\gtrsim 0.3$. This engenders from mild change in interaction free energy density of PNIPAm between volume fractions $\sim 0.4$ and $\sim 0.8$ (see \fref{Fint_PNIPAm}). In the narrow range of graft density in which volume fraction in a large section of a brush is in this range, on increasing graft density, brush accommodates extra monomers closer to grafting surface, incurring small interaction energy penalty to achieve only a small increase in stretching energy in the brush. For temperatures at which vertical phase separation is possible, $n(T,\bar{\rho}_g)$ approaches $0$ for a range of graft densities. This is because the phase separation interface propagates upwards towards the brush free surface with increasing graft density. Due to this, increasing amount of monomers are packed within the dense region close to the grafting surface and brush height does not change considerably with increasing graft density.

\end{appendix}

\bibliography{references}

\end{singlespace}

\end{document}